\documentclass{article}

\usepackage{arxiv}

\usepackage[utf8]{inputenc} 
\usepackage[T1]{fontenc}    
\usepackage[colorlinks=true,citecolor=blue]{hyperref}
\usepackage{url}            
\usepackage{booktabs, caption}       
\usepackage{nicefrac}       
\usepackage{microtype}      
\usepackage{graphicx}

\usepackage{doi}

\usepackage{eurosym}
\usepackage{bm}
\usepackage{amsmath}
\usepackage{amstext}
\usepackage{amsfonts,amssymb,graphics,xspace,endnotes}
\usepackage{rotating}
\usepackage{lscape}
\usepackage{color}
\usepackage{hhline}

\usepackage[comma]{natbib}

\usepackage{amssymb,amsmath,amsthm,amsbsy,color}

\usepackage{natbib}
\usepackage{graphicx}

\usepackage{epsfig,subfigure}
\usepackage{enumerate}
\usepackage{adjustbox}
\usepackage{units}
\usepackage{caption}
\usepackage{setspace,multirow,lscape,longtable,dcolumn}
\usepackage{fullpage}
\usepackage[table, xcdraw]{xcolor}
\usepackage[capposition=top]{floatrow}

\usepackage{threeparttable}
\usepackage{dcolumn}
\usepackage[normalem]{ulem}
\usepackage{arydshln}
\usepackage{array, makecell, rotating}
\usepackage[algoruled,boxed,lined]{algorithm2e}
\useunder{\uline}{\ul}{}

\newcommand{\bs}{\boldsymbol}

\newtheorem*{theorem*}{Theorem}

\graphicspath{ {figures/} }

\title{Managers versus Machines: Do Algorithms Replicate Human Intuition in Credit Ratings?}

\date{\today}       

\author{Matthew Harding \\
        Department of Economics and Department of Statistics,\\
        University of California - Irvine,\\
        3207 Social Science Plaza B,\\
        Irvine, CA 92697\\
        \And
        Gabriel F. R. Vasconcelos\\
        Bank of Communications BBM, \\
        34 Barão de Tefé Avenue 20th and 21st Floors\\
        Rio de Janeiro, RJ, Brazil, 20220-460 }

\renewcommand{\headeright}{~}
\renewcommand{\undertitle}{~}
\renewcommand{\shorttitle}{Managers vs Machines}

\begin{document}
\maketitle

\begin{abstract}

We use machine learning techniques to investigate whether it is possible to replicate the behavior of bank managers who assess the risk of commercial loans made by a large commercial US bank. Even though a typical bank already relies on an algorithmic scorecard process to evaluate risk, bank managers are given significant latitude in adjusting the risk score in order to account for other holistic factors based on their intuition and experience. We show that it is possible to find machine learning algorithms that can replicate the behavior of the bank managers. The input to the algorithms consists of a combination of standard financials and ``soft" information available to bank managers as part of the typical loan review process. We also document the presence of significant heterogeneity in the adjustment process that can be traced to differences across managers and industries. Our results highlight the effectiveness of machine learning based analytic approaches to banking and the potential challenges to high-skill jobs in the financial sector.

\end{abstract}

\keywords{credit risk, machine learning, random forest, gradient boosting, neural networks, scorecard}
\thanks{The authors wish to thank Sean Klein, Ph.D. for access to the data behind this work. All errors are our own. The authors are grateful to seminar participants at Hamburg University, Melbourne Business School, the University of California, Riverside, and Vienna University of Business and Economics for many helpful suggestions. We also thank Jonathan Hersh for insightful and valuable conversations.}

\clearpage

\doublespacing

\section{Introduction}
Financial institutions regularly employ a mix of algorithmic processes and human intuition in order to analyze firm financials and determine the risk profile of a borrower before deciding whether or not to extend a loan. A typical commercial lender relies on one or more internal credit risk models in addition to various indicators provided by third parties. Just as crucially though bank managers are given significant latitude in influencing the final outcome by changing the algorithmically generated risk score based on their interpretation of firm financials or ``soft" information revealed in the process of determining the suitability of the borrower. The aim of this paper is to investigate the extent to which machine learning techniques can replicate the human element and reach the same decisions as experienced loan managers. 

This paper focuses on machine learning and the role and limitations of algorithms in the commercial loan sector. One immediate layer of interpretation of our research is whether it is possible to improve the current credit risk rating systems for financial institutions. This can
lead to higher profitability through better identification of less risky borrowers and reduced overhead costs from streamlining the risk rating process. Machine learning in the context of credit risk modeling has many practical applications for financial institutions who profit from borrowers
successfully repaying loans. It is also perhaps one of the oldest and most established applications of machine learning in the financial sector. The widespread usage of internal credit risk models by financial institutions is a relatively well-established phenomenon attributed in a large part to the Basel II accords.\footnote{Established in 1975, The Basel Committee on
Banking Supervision is comprised of senior ranking representatives from supervising authorities or central banks of several nations including the United States \citep{engelmann2006basel}. The Basel Committee periodically releases recommendations on banking laws and regulations such as the Basel II accords which outlines the modern credit risk rating process. Initially published by the Basel Committee on Banking Supervision in June of 2004, the Basel II accords addresses three main banking issues as described by \cite{van2008credit}: (1) identifies the minimum capital requirements banks should exhibit in order to mitigate credit, market, and operational risk; (2) provides an overview of supervisory processes necessary to effectively assist banks in their assessment of risk factors; (3) describes the disclosure requirements for financial market agents to assess the risk and capital solvency of institutions. Within the second pillar of Basel II lies the foundation guidelines for the modern credit risk measurement system.} Developing quantitative methods to assess borrower default risk has been historically well studied, e.g.  see \citet{babii2020binary, lieli2010construction,  crook2004does,  fortowsky2001credit, martens2010credit}.

This study goes a step further however by considering the actual process through which credit risk ratings are being assigned and which in practice consists of a combination of algorithmic evaluations and subjective judgments by loan managers (see \citet{hand1997statistical} for a review on credit scoring methods). In a typical credit risk rating process, a borrower is
initially assigned a preliminary risk rating through a quantitative method, such as the \textit{scorecard approach}, which assesses the borrower’s likelihood of defaulting if issued a loan. Upon conclusion of the initial risk assessment, a loan manager reviews the borrower’s preliminary risk rating in conjunction with all available borrower information to determine the borrower’s final risk rating. We study this subjective adjustment of the borrower’s preliminary risk rating to their final risk rating commonly referred to as the rating adjustment process (or informally as ``notching"). The final risk rating is the credit risk rating used by the financial institution when making the decision to approve or deny the borrower’s request for a loan. In this setup, it is important to point out that the loan manager responsible for the rating is not the same as the loan officer, which is responsible for finding new deals to the bank. In order to minimize conflicts of interest, the first has his/her performance evaluated by how much risk is taken and the second by how many operations are made. This separation will induce officers to look for deals that are likely to be well rated by the managers, which are usually good deals aligned with the willingness of the bank to take risk. 

By shifting attention from the usual applications of machine learning, as part of the scorecard process, to an analysis of the loan officer’s subjective review of a borrower’s loan application, which seems to have been broadly overlooked, we also move the central focus of this study to the quantitative modeling of the loan officer's behavior in the adjustment process. In other words, our objective is to reproduce human behavior using machine learning methods in order to simulate the risk adjustment process. This has a number of novel implications.

First, we are able to contribute to the debate on the future of work \citep{autor2020work} in the context of the financial sector. While much of the debate on the rise of Artificial Intelligence (AI) and its potential job displacement impacts has focused on low skill repetitive jobs, relatively little is known about its potential impact on high skill managerial positions. Some of the recent debate has focused on the use of ``augmented analytics" broadly defined as a combination of human judgment and algorithmically generated insights as the most likely implementation of AI in various areas of business intelligence. This study however finds reasons to doubt that in many areas the human component cannot be reliably replicated even in highly complex environments such as those pertaining to the financial sector.

Second, we contribute to the small but growing literature that tackles the debate of the relative performance of ``man" vs ``machine" in a variety of tasks. \cite{kleinberg2018human} first introduced the issue in a different context by considering whether the bail classification decisions made by judges can be replicated and improved upon by simple machine learning techniques. They persuasively argued that this is indeed so and algorithms have continued to capture the attention of both policy makers and the public at large. In fact in the 2020 election, voters in California were asked to replace the bail system with an algorithmic process entirely. A parallel set of concerns emerges from the analysis of human and algorithmic decisions by evaluating the resulting differences as perceived biases (of humans or algorithms) and trying to draw conclusions about the behavioral failures of humans or the limitations of algorithms. 

This debate has recently extended to the financial sector. \cite{coleman2020man} compare the investment recommendations of traditional research analysts with those of robo-analysts and conclude that there are significant differences in the way algorithms accomplish the task. Robo-analysts are less optimistic and revise more frequently, but their buy recommendations can generate returns for investors. This provides additional evidence on the impact of technology on financial markets (see \cite{abis2017man} for evidence on the impact of quantitative mutual funds). \cite{zheng2020innovation} develops machine learning algorithms which can better screen low quality patents thereby improving both the generality of the awarded patents and the number of patent citations. They show that many patents granted by human examiners are likely to expire early and also have additional repercussions in terms of the financial performance of the firms receiving the patents. \cite{erel2018selecting} employ machine learning to predict director performance and potentially assist in the selection of directors for company boards. \cite{van2020man} rely on machine learning to predict firm earnings' expectations and compare them to those of human analysts. They show that humans display a series of biases which are associated with negative cross-section return predictability. \cite{hersh2020algo} partner with a large bank to predict loan performance but find that when presented with algorithmic based insights, most managers do not change their beliefs.

One feature of the current literature on ``man" vs ``machine" is worth highlighting. Not only does the current literature find significant improvements in the performance of algorithms over that of humans, but researchers choose to interpret the differences between the decisions of humans and algorithms as evidence of conscious or unconscious bias. This to mirrors the earlier economics literature which contrasted the engineering estimate of a task, such as an environmental cleanup, with the actual implementation by policy makers and interpreted discrepancies as evidence of human biases, or worse as evidence of discriminatory policies (e.g. \cite{hamilton1999calculating}). It is of course possible that humans optimize different objective functions than algorithms which are not well understood and it is thus not always very clear in what sense we can conclude that humans are biased in the absence of a welfare function we all agree to.

Our paper takes a different approach in that we use machine learning to replicate the \textit{actual} behavior of human loan officers without judging whether the decisions they make are biased or not. We think it is important to separate the question of replicating the behavior of humans from the question of improving on humans which naturally requires us to make a choice as to the criterion by which to judge the outcome of the process. In the context of loan origination this is particularly relevant since the bank may have a number of goals in addition to minimizing defaults e.g. a social goal such as assisting minority owned businesses. We leave this second question to future research in part because in the context of commercial lending defaults are extremely rare and loan agreements are typically renegotiated individually in according to complex rules and would necessitate a very different type of analysis.

We find that human managers assign final ratings which deviate substantially from the preliminary ratings assigned by the scorecard algorithm and that relatively simple machine learning algorithms can learn to replicate the choices made by human managers with a high degree of accuracy exceeding 95\%.  We show that financial variables play an dominant role in determining the final rating. On the one hand they are summarized by the scorecard risk rating and we show that even if we exclude that score from the analysis our machine learning algorithms will rely heavily on the financial variables, thus in essence replicating a version of the scorecard algorithm as part of the final risk rating assignment. But, human managers also adjust the ratings using other variables especially qualitative variables related to the managerial quality of the borrower but lower weight is put on these variables in the final risk determination.

We also document significant heterogeneity in the extent to which final risk ratings vary from the preliminary ones. The majority of the adjustments happen for loans with preliminary risk ratings in the middle of the risk distribution. We document that some of the heterogeneity can be traced to manager specific fixed effects. Risk considerations also appear to vary by industry. This indicates that a mixture of factors may be at work. Human managers use their specific skills and experience to compensate for the fact that the scorecard does not vary sufficiently with the individual characteristics of the loans in different industries, while also imposing a non-trivial amount of idiosyncratic discretion on the final rating assessments. We also show however that the amount of discretion appears to have diminished over time, which may indicate either improvements in the scorecard or increased reliance on automatic ratings over the past decade in the banking sector.

The paper is thus structured as follows. In Section 2 we introduce the data used in the analysis and discuss the process through which commercial loans are assigned risk ratings by algorithmic means. We then highlight the extent to which these ratings are manually adjusted by human managers to reflect their intuitions and experiences. In Section 3 we introduce a number of different machine learning techniques which are quite different in nature and we hypothesize that they may correspond to different models of information processing that may be employed by the human bank managers. In Section 4 we compare the performance of algorithms to the observed outcomes of the human managers and argue that some algorithms perform remarkably well. We show that algorithms do in fact approximate the risk adjustments performed by humans and explore in what way they utilize the available information. We also highlight the presence of heterogeneity in outcomes which is then explored further in Section 5 in relation to individual managers and across industries. Section 6 concludes.

\section{Data}
This paper uses a unique proprietary dataset compiled by a team of analysts at a large commercial bank for the purpose of this project. It consists of a portfolio of 37,449 rating events to 4,414 unique customers between January 2010 and December 2018. The customers are firms of various sizes and across diverse industries largely based on the West Coast of the United States. In this context each customer is a borrower and it is common for the same customer to hold multiple loans an lines of credit with the bank. Some customers are observed to have in excess of 50 rating events with the bank. For the purpose of conducting this research a number of different internal databases were linked by customer. These are typically maintained by separate teams within the bank. Some consist of information used in the scorecard. Other internal data however tracks a variety of additional financial information on each customer which may be used to adjust risk at a later stage, re-negotiate the terms of the loans or conduct stress testing. Since we do not know for sure which information is actually reviewed by the human managers we assume that all information available internally to a manager should be part of the predictive analysis. Our main interest is the process which assigns risk ratings to each loan. It is beneficial to review the details of this process in the next subsection. We will then introduce the variables recorded for each loan and customer and which are likely candidates for predicting the ratings.

\subsection{The Credit Risk Assessment Process}

The credit risk assessment process begins for a particular borrower with compiling data for analysis\footnote{Managers also have the power to change balance sheet information from the company. For example, banks may use their own evaluation of real state, which would impact the tangible assets of the company. This is a step of human intervention in the credit rating that affects also the outcome of the preliminary risk rating generated by the scorecard and we do not observe that in out data.}. The data collected will represent general information on the borrower and their credit history. Upon conclusion of data gathering, a borrower’s information is input into a financial institution’s proprietary model that outputs a preliminary rating proposal. Typically this consists of a \textit{scorecard}. A credit scorecard seeks to identify a borrower’s probability of defaulting based on the borrower’s characteristics. The definition of what constitutes a default is left to the discretion of the particular financial institution. The characteristics selected within the scorecard have been statistically determined to be significant indicators in identifying a borrower’s default status and the same characteristics are reviewed for every borrower.

In this paper we take both the scorecard algorithm and its output as given. Historically the scorecard is one of the earliest applications of machine learning to the financial sector. It consists of a model which selects characteristics and their attribute weightings, referred to as key characteristics. These key characteristics are input into a statistical model, (often as simple as a logistic regression) to map the attributes of the selected characteristics to the likelihood of defaulting. Once the statistical model is developed, the results are synthesized into a viable scorecard.

Returning to the credit risk assessment process, a borrower’s characteristics are input into the scorecard to which the borrower is given a sum total score of the point values associated with their attributes for the characteristics. The borrowers sum total score from the scorecard is then mapped to an ordered risk rating to serve as the borrower’s preliminary risk rating. Once a borrower has been assigned a preliminary risk rating, a loan officer reviews the rating and other information either included or withheld from the scorecard process to determine the borrower’s final risk rating. The final risk rating, determined by the loan officer, is the borrower’s credit risk assessment used in underwriting decisions for loan products and other bank products.

It is this last phase of the credit risk assessment process, when a borrower’s preliminary risk rating is adjusted by a loan officer, that is the focus of this paper. Most research has been directed at the quantitative approach to assigning a borrower’s preliminary risk rating through a scorecard and other methods, yet limited research has been conducted to better understand the loan officer’s rating adjustment behavior. 

Further, this paper is not concerned with evaluating the accuracy of the ratings or suggesting alternative rating processes. In fact we shall ignore whether or not loans default and focus on modeling the loan officers' adjustment process behavior as observed. This is partly due to data limitation and also issues related to how defaults are processed in the commercial loan industry. Even though we have a very large portfolio of loans available for the analysis we only observe 359 defaults in the data which corresponds to a default rate of less than 1\%. This is explained by a number of factors including the common practice in this industry of frequently renegotiating the terms of a loan along complex dimensions that are beyond the scope of this paper.

Recall that our objective is to model the as of yet unknown factors that the human underwriter considers when assigning risk rating adjustments. This is captured by considering the two ratings separately. The first rating is the scorecard rating which is generated solely through the scorecard process. It consists of ratings on a scale from 1 to 15 where 1 represents a low risk of default and 15 represents a relatively high risk of default. Note that for the scorecard data observed by us, no customer received a risk rating of 1 in the data set. A manager then reviews the available information on the customer which typically goes beyond the information utilized in the scorecard. It is expected that this process includes third party risk assessments, additional financial information and reviews of soft information such as the managerial quality of the leadership team of the borrower. The manager then assigns a final risk rating which is allowed to differ from the preliminary rating generated by the scorecard algorithm.

\begin{figure}[ht]
\caption{Empirical distribution of ratings and ``notches" of the loan portfolio.}
\label{f:hist}
\includegraphics[width=0.55\textwidth]{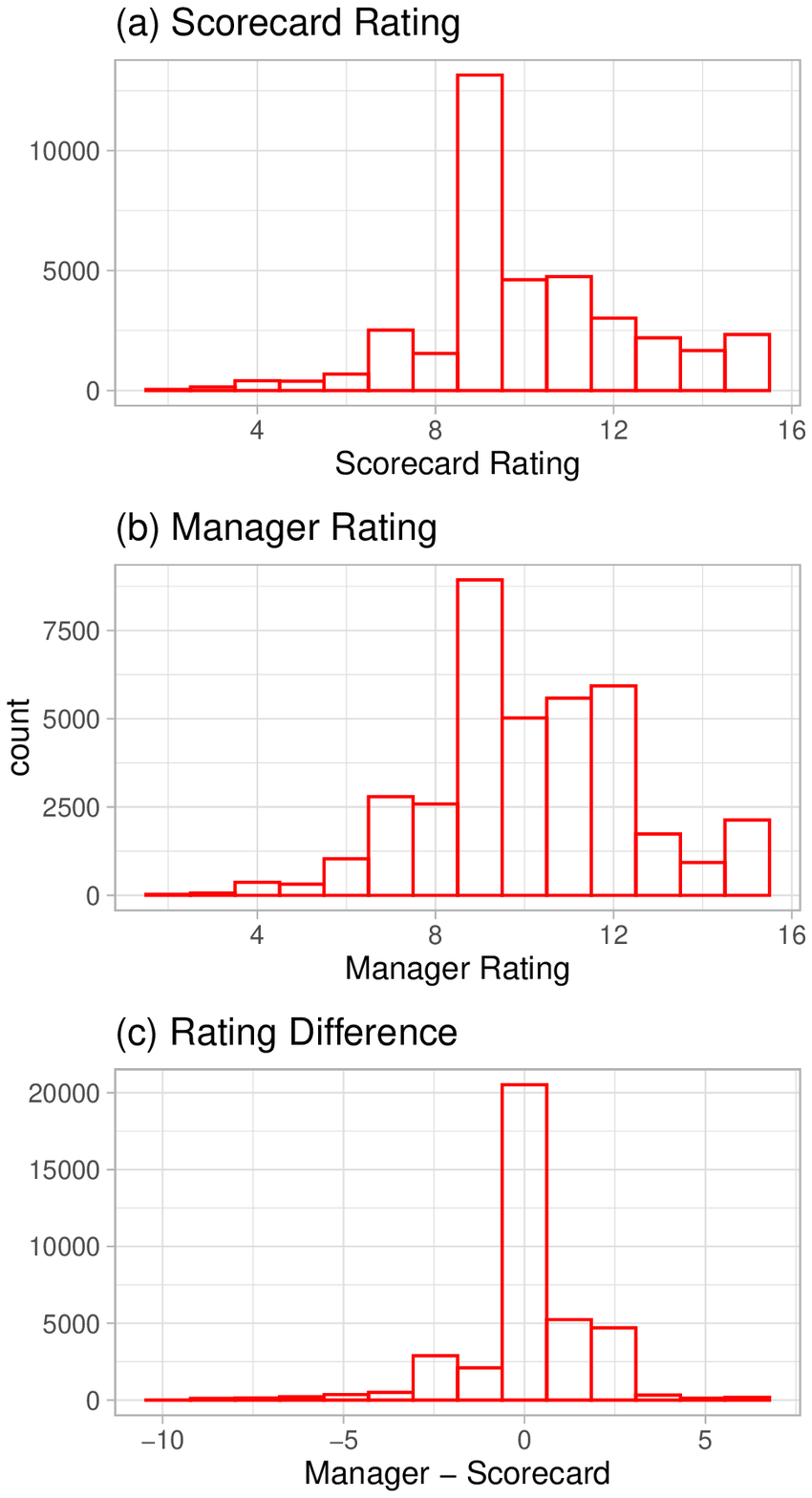}
\end{figure}

Figure \ref{f:hist} illustrates the outcome of the two steps of the rating procedure. In panel (a) we have the distribution resulting from the quantitative scorecard rating. In panel (b) we have the manager rating and panel (c) shows the difference between panels (b) and (a). The process of changing the risk ratings from those automatically generated by the scorecard is called \textit{``notching"}. The distribution of the notches in panel (c) shows that the majority of the notching is upwards. This corresponds to managers revising the riskiness of the loan upwards, i.e. 
managers are more likely to evaluate loans as riskier than the scorecard does.  

\begin{center}
[Figure \ref{f:hist} here]
\end{center}

An alternative way of visualizing the ratings and changes in the same diagram corresponds to displaying the confusion matrix in Table \ref{tab:cms} which presents both the scorecard and the manager ratings in the same table. Each cell contains the corresponding number of loans for a given combination of scorecard and manager ratings. Thus, the diagonal of the table counts the number of loans for which the loan manager does not change the rating generated by the scorecard. The off-diagonal cell count the loans which are ``notched" either up or down. Cells below the diagonal correspond to loans which are considered riskier by the human manager than the scorecard algorithm while cells above the diagonal correspond to loans considered less risky by the manager than the scorecard algorithm. We also indicate the relative density of the cells by additional shading for ease of reference. 

\begin{center}
[Table \ref{tab:cms} here]
\end{center}

Some features of this confusion matrix are worth noting to establish some empirical facts about the notching process. First, managers deviate from the scorecard frequently and some deviations are significant. For example there are cases where a rating of 7 has been updated to a rating of 13 by the manager. Second, there are more loans below the diagonal of the table than above, which means that managers are more likely to classify loans as riskier than the scorecard. Rating managers and loan officers are different individuals, the first group is responsible to control how much risk is taken by the bank and the second group is responsible for getting new clients and new deals. This setup explains why managers classify loans as riskier than the scorecard and it avoids conflicts of interest because risk managers will be aligned with the bank's willingness to take risk and officers will look for deals that are more likely to obtain a good rating from the risk managers. Third, the majority of the loans that are notched are in the middle range of the risk distribution. This is partly a result of the overall distribution of risk ratings in the portfolio, where we see relatively few loans that are rated as having very low risk. There are more loans which are rated as being high risk but the most common risk rating in the data is 9. Loans in that group are also most likely to be adjusted by 1-2 risk categories. For example 7,671 loans with a scorecard rating of 9 are not adjusted while 3,133 loans are notched to a risk rating of either 10 or 11. Likewise, 1,304 loans are notched in the reverse direction to either 8 or 7.

\subsection{Predictors}

One of the attractive features of our data consists in the bank's willingness to compile and share with us a remarkable range of variables on each individual loan. In addition to the two risk ratings already discussed we will rely on a large number of variables to create predictive models of the ``notching". The variables are listed in Table \ref{tab:variables} and briefly defined below.

To provide structure to these other variables, we can classify them into three categories: quantitative variables, categorical variables and qualitative variables. Quantitative variables correspond to financial variables.  The financial variables represent the financial information for customers at the point in time when the rating is decided. This information is extracted from the company’s balance sheet, income statement, and cash flow statement. Typical variables include net sales, total assets, or the ratio of total debt to market capitalization. Categorical variables capture further discrete objective characteristics of the customer. For example this may include the industry in which the customer operates or further details about the loan such as whether it is syndicated or not. 

A very interesting category of variable consists of qualitative variables which attempt to capture subjective classes or evaluations. For example one variable captures the assessment of the client's management quality by answering the question: ``How has the current management team/owners responded to adverse conditions or economic downturn ?". The answer is then coded into different categories. For example a B rating is entered when the client "Forewarned the Bank, was responsive and did not take significant steps to cure the situation. Responsive management that proactively identified the problem(s), informed the Bank when appropriate and legally permissible, and cooperated with the Bank throughout the process; and Management’s actions were not decisive, and their minimal action might also have exacerbated the problem or possibly proved successful, but followed a different strategy than Bank’s preference." 

\begin{itemize}
    \item Rating Variables:
    \begin{itemize}
        \item \textbf{prelim\_risk\_rating:} Preliminary risk rating calculated from the scorecard used by the managers as input for the final risk rating. It is rated on the scale between 1 (very save) to 15 (very risky). 
        \item \textbf{final\_risk\_rating:} Risk rating calculated by managers in the same 1-15 scale. 
    \end{itemize}
    \item Quantitative Variables:
    \begin{itemize}
        \item \textbf{net\_profit:} The amount by which income from sales is larger than all expenditure. Also called profit after tax.
        \item \textbf{net\_sales:}  The sum of a company's gross sales minus its returns, allowances, and discounts. Revenues reported on the income statement often represent net sales.
        \item\textbf{net\_profit\_margin: The percentage of revenue left after all expenses have been deducted from sales.}
        \item \textbf{cashoprofit\_to\_sales:} Gross cash flow from all operating activities divide by sales. 
        \item \textbf{total\_assets:} This defines total assets in detailed balance sheet.
        \item \textbf{total\_liab\_by\_tang\_net\_worth:} The ratio between total liabilities and tangible net worth.
        \item \textbf{acf:} Recurring cash flow available to debt service.
        \item \textbf{total\_debt\_to\_cap:} Total debt divided by market capitalization.
        \item \textbf{total\_debt\_to\_acf:} Total debt divided by acf.
        \item \textbf{end\_cash\_equv\_by\_tot\_liab:} The ratio between end cash equivalent and total liabilities. 
    \end{itemize}
    \item Qualitative Variables:
    \begin{itemize}
        \item \textbf{access\_outside\_capital:} Assessment of a firm's access to outside sources of funding market. Values under this variable include A (good), B, or C (bad).
        \item \textbf{level\_waiver\_covenant\_mod:} Level of waivers or modification of covenants. Values from A to D, where A is minor/none, B is moderate, C is significant and D is none/new borrower.
        \item \textbf{management\_quality:} Rating of the management quality from A to F, where A is the best possible, D is the worst possible, E is for companies that never experienced any financial distress and F is unknown. 
        \item \textbf{market\_outlook\_of\_borrower:} Assessment of the borrower's outlook in the market. Values of variable are: A (good), B, or C (bad).
        \item \textbf{mgmt\_resp\_adverse\_conditons:} Assessment of the management's ability to respond to adverse conditions. Rated in the same categories as management\_quality.
        \item \textbf{strength\_sor\_prevent\_default:} Strength of secondary source of repayment. A for strong, B for moderate and C for weak. 
        \item \textbf{vulnerability\_to\_changes:} Assessment of the firm's vulnerability to changes in the economic or regulatory environment. A for stable, B for average and C for unstable. 
    \end{itemize}
    \item Categorical Variables:
    \begin{itemize}
        \item \textbf{primary\_exposure\_type:} Current structure of the exposure (e.g. line of credit or term loan, lease, standby letter of credit).
        \item \textbf{pct\_rev\_3\_large\_cus:} Revenue from largest customers. This field is to answer the question what percentage of the borrower’s revenues come from its largest customer? Then, answer “A” means >40\%, while answer “B” means 20-40\%, and “C” stands for <20\%.
        \item \textbf{public\_private\_cd:} Identify public or private firms.
        \item \textbf{snc\_indicator:} Syndicated Indicator.
        \item \textbf{not\_for\_profit\_ind:} Indicates if a firm is not-for-profit (Y/N).
        \item \textbf{revolving\_ind:} Revolving loan indicator (Y/N).
        \item \textbf{lgd\_secured\_or\_unsecured:} Whether the exposure is secured by specific collateral, secured by blanket lien, secured by equity commitments or capital calls, solid waste management, secured by general pledge of revenues or unsecured.
        \item \textbf{industry}: 2 digits NAICS code of company industry.
    \end{itemize}
\end{itemize}

\section{Methods}

From a machine learning perspective we are dealing with a fairly standard classification problem. 
Loans need to be classified into 15 different risk categories where 1 is low risk and 15 is high risk. The statistical setup is as follows. Let $\bs{x_i} = (x_{i,1},\dots,x_{i,m})$ be an independently and identically distributed (IID) random vector of predictors and $y_i$ be a discrete response variable defined in classes. The predicted class $\hat{y}_i$ is determined by a general function $f(\cdot)$ and a set of parameters $\bs{\theta}$ that will depend on the chosen model: 

\begin{equation}
\hat{y}_i = f(\bs{x_i},\bs{\hat{\theta}})
\end{equation}

We have a number of models at our disposal. In principle it does not matter which model is employed as long as we are able to predict the final risk ratings as close to the ratings of the human managers as possible. We employ four different methods in our analysis. The first is the Multinomial Logistic LASSO (MNL-LASSO), which is a penalized multinomial logistic model and will be our benchmark. Second we implement a neural network and finally we employ two models based on Classification and Regression Trees (CART), Random Forest and Gradient Boosting.

\subsection{Multinomial Logistic LASSO}

Since the final risk rating in our data is a multi-level categorical variable, the benchmark approach to predicting the final risk rating is to use a multinomial logistic regression (MNL). Our MNL will predict a customer’s final risk rating using a customer’s preliminary risk rating, and other predictor variables (encompassed into the vector $\bs{x_i}$). The basic structure of the MNL with alternative-invariant regressors, can be thought of as predicting the probability customer $i$ belongs to final risk rating class $j$:

\begin{equation}
    p_{ij} = P(y_i = j) = \frac{exp\{\boldsymbol{x}'_i \boldsymbol{\beta}_j\}}{1
+ \sum^{14}_{k=2}exp\{\boldsymbol{x}'_i \boldsymbol{\beta}_k\}} \quad \textrm{for}
\quad j=2,...,15
\end{equation}

Additionally, we will treat $\boldsymbol{\beta}_5$ as a vector of zeros to normalize the parameters such that sum of the probabilities for a customer belonging to each class is equal to one. To add an extra layer of variable selection and to potentially improve the predictive power of the multinomial approach, we estimate the MNL with a LASSO regularization.\footnote{In earlier versions of the results we also reported the elastic net regularization but all attempts to predict the ratings using a MNL specification were significantly outperformed by other approaches.} 

The addition of a LASSO penalty in these models can be easily implemented my modifying the likelihood:\footnote{For additional details and computational considerations see  \citet{krishnapuram2005sparse}.}

\begin{equation}
    l(\bs{\beta}, \lambda, \alpha) = \frac{1}{N}\sum^N_{i=1} \sum^{15}_{j=2} y_{ij} ln(p_{ij}) + \lambda |\bs{\beta}|.
\end{equation}

The lambda hyperparameter value was selected by whichever lambda value produced the lowest average Mean Squared Error obtained through a 5-fold cross-validation procedure. 

Note that we assume that each rating can be chosen by the model for any given customer observation conditional on the observables. This model neglects a rank order applied to the final risk rating when it determines a customer’s probability of belonging to a specific rating class. Essentially, we will assume for the MNL that a customer can be classified with a final risk rating of 2 or of 15. In theory, this assumption should contradict with the real behavior of a loan officer because we want a sense of stability in the risk rating adjustment process. While it is true that a customer with any given preliminary risk rating can be notched to any given final risk rating, customers with preliminary risk ratings toward the extreme values are generally adjusted within a tighter window than a customer with a less extreme risk rating value. This observation can be witnessed when reviewing the empirical distribution of rating adjustments. The inequality in probability assignment among the range of final risk ratings may be one of the reasons why tree based models are likely to outperform multinomial approaches.

\subsection{Neural Networks}

Neural Networks are extremely popular and considered by many to be the most powerful machine learning techniques currently available due to their attractive universal approximator property for any smooth prediction function (see \citet{hornik1989multilayer} and \citet{cybenko1989approximation}). Neural Networks are very flexible on its architecture to adjust to different types of data using different setups of hidden layers, neurons and activation functions. These complex networks are also known as "deep learning", which is capable of modeling very complex nonlinear interactions between predictors.\citep{farrell2020deep} 

We implement a "feed forward" neural network as in  \citet{medeiros2019forecasting} and \citet{gu2018empirical}. It is a model that consists of an input layer of raw predictors, one or more hidden layers and an output layers that that aggregates the results from the hidden layer into the outcome prediction. The input layer has the same number of elements as the number of predictors, which we will define as $k_0$. The signal is sent from the input layer to the first hidden layer and it is transformed according to a $k_0+1$ parameters vector $\theta_{i,j}$ that includes one intercept and one weight for each predictor. Here $i$ indexes the neuron and $j$ indexes the hidden layer. An activation function $f(\cdot)$ is applied to the final result and a signal is sent to the next hidden layer or the output layer. For example, suppose we are starting in the input layer and we are looking at the third neuron of the first hidden layer ($j = 1$ and $i =3$). The input layer is indexed as $j = 0$ and $x_0$ are the raw predictors so that $x_j$ are the predictors resulted from the layer $j$. Equation (\ref{eq:nn1}) shows the result for this example. 

\begin{equation}
\label{eq:nn1}
    g(x_{0},\theta^{3,1}) = f\left(\theta_0^{(3,1)} + \sum_{z = 1}^{k_{0}} x_{z,{0}} \theta_z^{(3,1)}\right)
\end{equation}

Equation \ref{eq:nn1} can be generalized into equation \ref{eq:nn2}, which shows the results starting from a general layer $j-1$ to the next layer $j$ and a general neuron $i$.  

\begin{equation}
\label{eq:nn2}
    g(x_{j-1},\theta^{i,j}) = f\left(\theta_0^{(i,j)} + \sum_{z = 1}^{k_{j-1}} x_{z,{j-1}} \theta_z^{(i,j)}\right)
\end{equation}

The neural network employed has three hidden layers with 56, 42 and 28 neurons respectively and ReLU activation function (equation (\ref{eq:relu})) and one final layer with Sofmax (\ref{eq:soft}) activation function and the dimension equal to the number of classes in the model. The output layer will be a vector where the dimension is also the number of classes and the values represent the probability of a particular observation to belong to a particular class. 

\begin{equation}
\label{eq:relu}
    f(x) = \max(0,x)
\end{equation}

\begin{equation}
\label{eq:soft}
    f(x_i) = \frac{e^{x_i}}{\sum_j e^{x_j}}
\end{equation}

Neural Networks estimations were made using the \textit{Keras} interface for R \citep{keras}.  

\subsection{Classification and Regression Trees}

Classification and Regression Trees (CART) \citep{breiman2017classification} are the building block of Random Forests and Gradient Boosting. A CART is a nonparametric model that uses recursive partitioning of the covariate space to approximate an unknown nonlinear function with local predictions . In other words, it aims to partition the space of predictor variables with partitions that are orthogonal to the axis of the variables. Following the notation in \citet{fonseca2018boost}, a classification tree model with $K$ terminal nodes (leaves) approximates the function $f(\bs{x_i},\bs{\hat{\theta}})$ with an additive model $H(\bs{x}_i;\bs{\psi})$ and a set of parameters $\bs{\psi}$. $H(\cdot;\cdot)$ is a piece-wise constant function with $K$ subregions that are orthogonal to the axis of the predictor variables. Each subregion represents one terminal node. The tree also has $J$ parent nodes. 

Any tree can be represented by the following notation. The root node is at position 0. Each parent node at position $j$ is split into two child nodes at position $2j+1$ and $2j+2$. Each parent node has a split (threshold) associated with a variable, $x_{s_j,i}\in\bs{x}_i$, where $s_j\in\mathbb{S}=\{1,2,\ldots,m\}$. The sets of parent and terminal nodes are represented respectively by $\mathbb{J}$ and $\mathbb{T}$ and they uniquely identify the architecture (structure) of the tree.

Therefore,
\begin{equation}\label{E:model}
H(\bs{x}_i):=H_{\mathbb{J}\mathbb{T}}(\bs{x}_i;\bs{\psi})=\sum_{k\in\mathbb{T}}\beta_kB_{\mathbb{J}k}\left(\bs{x}_i;\bs{\theta}_k\right),
\end{equation}
where $0\leq B_{\mathbb{J}k}\left(\bs{x}_i;\bs{\theta}_k\right)\leq 1$ is a product of indicator functions defined as
\begin{equation}\label{E:B}
B_{\mathbb{J}k}\left(\bs{x}_i;\bs{\theta}_k\right)=
\prod_{j\in\mathbb{J}}I(x_{s_j,i};c_j)^{\frac{n_{kj}(1+n_{kj})}{2}}
\left[
1-I(x_{s_j,i};c_j)
\right]^{(1-n_{kj})(1+n_{kj})},
\end{equation}
with
\begin{equation}
I(x_{s_j,i};c_j)=
\begin{cases}
1 & \textnormal{if}\, x_{s_j,i}\leq c_j (continuous),  x_{s_j,i} \in c_j (discrete)\\
0 & \textnormal{otherwise},
\end{cases}
\end{equation}
and
\begin{equation*}
n_{kj}=
\begin{cases}
-1 & \text{if the path to leaf } \,k\, \text{ does not include the parent node } j; \\
0  & \text{if the path to leaf } \,k\, \text{ include the right-hand child of the parent node } j; \\
1  & \text{if the path to leaf } \,k\, \text{ include the left-hand child of the parent node } j. \\
\end{cases}
\end{equation*}
where $c_j$ is the splitting point from a split in variable $j$ or it indicates the classes that will go to each child node if $j$ is discrete. Equation (\ref{E:B}) will use the combination of 0 or 1 exponents to draw the path to each terminal node in the tree. Furthermore, define $\mathbb{J}_k$ as the set of indexes of parent nodes included in the path to leaf (region) $k$, such that $\bs{\theta}_k=\{c_j\}$ with $j\in\mathbb{J}_k$, $k\in\mathbb{T}$. Finally, it is clear that $\sum_{k\in\mathbb{T}}B_{\mathbb{J}k}\left(\bs{x}_i;\bs{\theta}_k\right)=1$.

To grow a CART model one must follow an iterative process that starts in the root node and expands the tree until the terminal nodes. Each new step is equivalent to a new split in some node. Each split includes a step to determine which variable $j$ and splitting point $c_j$ will be used. The best split will be the one that minimizes a loss function such as the Gini index or the a quadratic loss. The tree procedure stops when a criterion such as the minimum number of observations per terminal node or a maximum depth is reached. 

Algorithms such as Gradient Boosting and Random Forests make use of a combination of CART models in order to achieve a more stable model since a small change in the data may cause significant changes in a single CART model because of its iterative architecture. We will discuss these ensemble algorithms in the next sections. 

\subsection{Random Forests}

Random Forests were originally proposed by \cite{breiman2001random}. Once we have defined CART models, the transition to Random Forests becomes relatively simple. They are bootstrap aggregating models applied on CART with some minor adjustments to ensure that one model is as different as possible from the other. Bootstrap aggregating \citep{breiman1996bagging} (Bagging) consists of estimating the same model $B$ times on bootstrap samples with replacement, usually the same size as the original sample. The main adjustment made to CART estimation is to look only at a random subset of all variables to make each split instead of looking at all variables. This ensures that the difference between trees is bigger, which benefits bagging estimates. The procedure is as follows:

\begin{itemize}
    \item (1) Draw a boostrap sample $s_b$ of the same size as the data with replacement.
    \item (2) Grow a classification tree on $s_b$ where in each new step we must select $k*<K$ variables to check for the best new split. 
    \item (3) Repeat steps (1) and (2) $B$ times. In a classification model, the outcome of each tree will count as a vote. The final prediction for each observation will be the class that had more votes. 
\end{itemize}

We used the package \textit{ranger} \citep{ranger} in R to estimate the Random Forests. As for the parametrization,  the number of candidate variables for each split was the square root of the total number of variables, chosen randomly in each split. The tree size was controlled by the minimum number of observations allowed in each terminal node, which was 5 and the number of trees in the Random Forest was 500.

\subsection{Gradient Boosting}

Gradient Boosting was proposed by \citep{friedman2001greedy}. It is a greedy method to approximate nonlinear functions that uses base learners for a sequential approximation. In a CART context, instead of growing independent trees like the Random Forest, the Gradient Boosting grows the next tree in the pseudo residuals of the previous tree. 

The algorithm initializes with some naive prediction, for example, in step 0 make $\phi_{i,0} = c_{max}$, where $c_{max}$ is the most common class in the sample and the second. The next step (step 1) is to calculate the pseudo residuals for a given loss function $L$, defined as:

\begin{equation}
    \tilde{u}_{i,1} =  \frac{\partial L(y_i,\phi_{i,0})}{\partial \phi_{i,0}}
\end{equation}

Next we grow a CART $u_{i,1} = H_1(\bs{x}_i;\bs{\psi_1})$ on $u_{i,1}$ and a set of control variables $\textbf{x}_i$. The subscript on $H$ indicates the step of the algorithm. The following step is the gradient step where we will define how the new information from $H$ enters in the additive model. The goal is to fin $\rho_1$ such that

\begin{equation}
    \rho_1 = \text{arg min}_\rho L(y_i, \phi_{i,0}+ \rho \hat{u}_{i,1}  ).
\end{equation}

Finally we update the previous $\phi_0$

\begin{equation}
\label{equpdate}
    \phi_{i,1} = \phi_{i,0} + \rho_1 \hat{u}_{i,1}
\end{equation}

The procedure above accounts for one iteration of the Gradient Boosting. The next iteration starts by going back to the first step using $\hat{\phi}_1$ instead of $\hat{\phi}_0$. Shrinkage can be applied to the algorithm in order to avoid over fitting. However the trade-off is that one will need more iterations until convergence. The most usual way to apply shrinkage is to modify equation \ref{equpdate}

\begin{equation}
    \phi_{i,1} = \phi_{i,0} + v\rho_1 \hat{u}_{i,1},
\end{equation}

where $v \in (0,1)$. As result, the model will use less information from each step and it will be less dependent of the fit of individual CART models. Smaller values of $v$ make convergence slower, but the model more robust to over-fitting. We used $v=0.05$ and  the number of iterations $M=1000$. We also used the same procedure as the Random Forest of randomly selecting a subset of variables in each new split of the trees, in this case, the proportion was $2/3$. The loss function was the Softmax. All estimations of the Gradient Boosting were made with the \textit{xgboost} \citep{xgboost} package in R.

\begin{algorithm}
 \caption{Gradient Boosting}
 \label{A:2}
 initialization: For a loss function $L$ set the first step $\phi_{i,0}$ \;
 \For{m=1,\dots,M}{
  make $\tilde{u}_{i,m} =  \frac{\partial L(y_i,\phi_{i,m-1})}{\partial \phi_{i,m-1}}$\;
  grow a CART to fit $u_{im}$ \;
  make $\rho_m = \text{arg min}_\rho L(y_i, \phi_{i,m-1}+ \rho \hat{u}_{i,m}  )$\;
  update $\phi_{i,m} = \phi_{i,m-1} + v\rho_m \hat{u}_{i,m}$ where $v\in(0,1)$
 }
\end{algorithm}

Algorithm \ref{A:2} generalizes the Gradient Boosting procedure to $M$ iterations. More details on how the algorithm works can be found for example in \cite{friedman2001greedy}, \cite{zhang2005boosting} and \cite{buehlmann2006boosting}. 

Gradient Boosting has been very successful in several fields both in the industry and academia especially after the \citep{xgboost}, which has a very efficient and flexible implementation. Examples in finance are \citep{carmona2019predicting}, \citep{chang2018application}, \citep{krauss2017deep} and  to mention a few.

\subsection{Behavioral Interpretation of Machine Learning Models}

It might be worth pausing for a moment to consider the interpretations of these different models before reviewing the estimation results. Each of these models processes information in quite distinctive ways.\footnote{We do not discuss simpler models such as logit or linear regression due to their extremely poor performance in this setting.} While ultimately machine learning models are trying to approximate an unknown function $f(.)$, their approach to incorporating the input data is quite different. To the extent that we are willing to consider each algorithm as a model of human cognition they reflect very different approaches to transforming the available input data into a final rating. The penalized multinomial logistic model is an algorithm where the input data enters as a linear index. The penalization however selects among the input variables and the final rating is only a function of a limited number of variables. One might reasonably consider this to be a model of behavioral inattention \citep{gabaix2019behavioral}. Neural networks by contrast are designed to utilize all input variables and combine them into highly nonlinear functions. Decision trees are particularly effective at partitioning the data and selecting among interactions between variables. Implementations such as random forests average predictions based on several models which can be thought of as related but distinct scenarios that a manager may use to evaluate the risk of a particular customer. While we do not wish to argue that there is an exact correspondence between the machine learning models and any given model of the human mind, it is also worth considering whether the features driving the success or failure of a particular algorithm in the data may not be indicative of the way humans process information too.

\section{Empirical Findings}

Let us first review the performance of the four different machine learning techniques in predicting the final risk rating\footnote{There are two potential variables we could predict, the final risk rating and the change between the final and the preliminary risk ratings. This definition has a very small impact on the predictions because predicting the right final rating is equivalent to predicting the right change. However, the interpretation of the impact of the variables in the predicted outcome is different in these two cases. We adopted the final risk rating as our target variable because it has the same meaning as the preliminary rating and it is ultimately what we are trying to predict.} for each loan. In Table \ref{tab:cv} we report several metrics on the prediction results. There are a number of different metrics that could be employed to evaluate the performance of the methods. We also estimate the various models in two different setups, one with the scorecard rating as a predictor (top panel of the table) and the other without it (bottom panel of the table). The most familiar metric is the Root Mean Square Error (RMSE). The Accuracy measures the proportion of correctly classified loans. RMSE is more sensitive to large deviations of the predicted value from the true value. We also report a 95\% confidence interval for the Accuracy which is computed based on a standard binomial test (AccuracyLower, AccuracyUpper). As is typical in machine learning we also report the Accuracy p-value computed from a one-sided test \citep{kuhn2008building}  which compares the prediction accuracy to the "no information rate", which is the largest class percentage in the data (23.85\%).

\begin{center}
[Table \ref{tab:cv} here]
\end{center}

The results indicate that all machine learning models are superior to the no information rate. But we also find substantial differences in the performance of these models. The poorest performance is observed for the MNL LASSO both in terms of Accuracy and RMSE. In fact it classifies just over 53\% of the loans in the same risk category as the manager. By contrast the other three methods classify over 90\% of the loans in the same risk category as the manager. The best performance is observed for the Random Forest algorithm which classifies just under 96\% of the loans correctly. The Boosting algorithm is a very close second with a similar Accuracy and slightly lower RMSE. The Neural Network classifies 93\% of the loans correctly but has a RMSE which is about 37\% higher than that of the Random Forest. This indicates that when the Neural Network makes mistakes it tends to classify the loans into risk categories that are several notches away from the rating that the human manager chose. The Accuracy confidence intervals for the Random Forest and for Boosting overlap but are distinct from those for the Neural Network. As previously discussed these results don't indicate that one method outperforms the other one as far as an external metric of accuracy is concerned such as the objective default probability. It is quite possible that the Neural Network for example is best for that classification task. Here the objective is only to replicate the choice of the human manager and for this task the Random Forest seems to outperform all other methods across the metrics investigated.

In the lower panel of Table \ref{tab:cv} we re-computed the same models but excluded the preliminary risk rating based on the internal scorecard algorithm from the set of predictor variables. As expected, the performance of all algorithms diminishes. For the Random Forest model the Accuracy decreases by about 1.3\% but the RMSE increases by almost 12\%. Also, the confidence interval of models with and without the preliminary rating does not overlap. The performance loss is larger for the Neural Network, 3\% decrease in terms of Accuracy, and a 15\% increase in the RMSE. This indicates that even though the same information used to construct the scorecard is available to the machine learning algorithms they are not able to replicate the scorecard perfectly. This seems to affect the probability of assigning risk categories that differ by several notches from the manger assigned risk rating. While we do not know the proprietary scorecard algorithm employed, it utilizes pre-trained weights derived from a large number of datasets. There are of course many other varieties of deep learning architectures that could be constructed, it is worth noting that a fairly standard approach, such as the one implemented here, fails to outperform tree based methods and furthermore it does not replicate the scorecard output to a high degree of accuracy either. 

\begin{center}
[Table \ref{tab:cmrf_full} and Table \ref{tab:cmrf_noscore} here]
\end{center}

In order to explore the performance of the Random Forest models further, we also present the full confusion matrices for the model with and without the scorecard risk rating in Table \ref{tab:cmrf_full} and Table \ref{tab:cmrf_noscore} respectively. In contrast to the confusion table discussed earlier which compared the scorecard rating with the final manager rating, in these tables we compare the manager's final rating with the model predicted final rating. Both tables have a very small number of loans concentrated outside of the diagonal and large misclassification errors between classes do not happen. The biggest confusion is between classes 9 and 10 and 10 and 11 and the errors are not concentrated in one direction. Both matrices have similar structure but the counts outside the diagonal are smaller in Table \ref{tab:cmrf_full} because it is more accurate given the inclusion of the scorecard as a predictor. The concentration of the errors in the middle range of the rating distribution may be indicative of other factors that drive the rating when it is neither high nor low. This may indicate further unobserved heterogeneity which we explore in the next section.

Overall, the performance of the machine learning models is surprisingly good at reproducing human behavior with an Accuracy in excess of  95\%. The good performance of Random Forests and the somewhat disappointing performance of Neural Networks is broadly consistent with the evidence from recent econometrics studies. In contrast to typical machine learning applications such as image processing or natural language processing, financial data is inherently noisy and presents additional challenges that are not well understood in the computer science literature at the moment. Random Forests, while a much simpler tool from a machine learning perspective, tends to perform much better in finance and economics since it provides both a variable selection mechanism and a reliance on an algorithmic exploration of the interactions between variables. Both features have been fundamental to economics and financial modeling for decades. From a practical perspective it is worth noting that our results may indicate that the bank could process loans faster and cheaper in the absence of human loan managers with very comparable results. While managers naturally perform a variety of tasks, it is hard to argue that they are essential for this particular task and a relatively simple algorithm can perform just as well. It is also important to note that with additional data and computational power these algorithms can be further improved as well.

\subsection{Variable Importance}

\begin{figure}[htb]
\caption{Variable Importance for the Random Forest with the Scorecard in the Model}
\label{f:importance_full}
\includegraphics[width=1\textwidth]{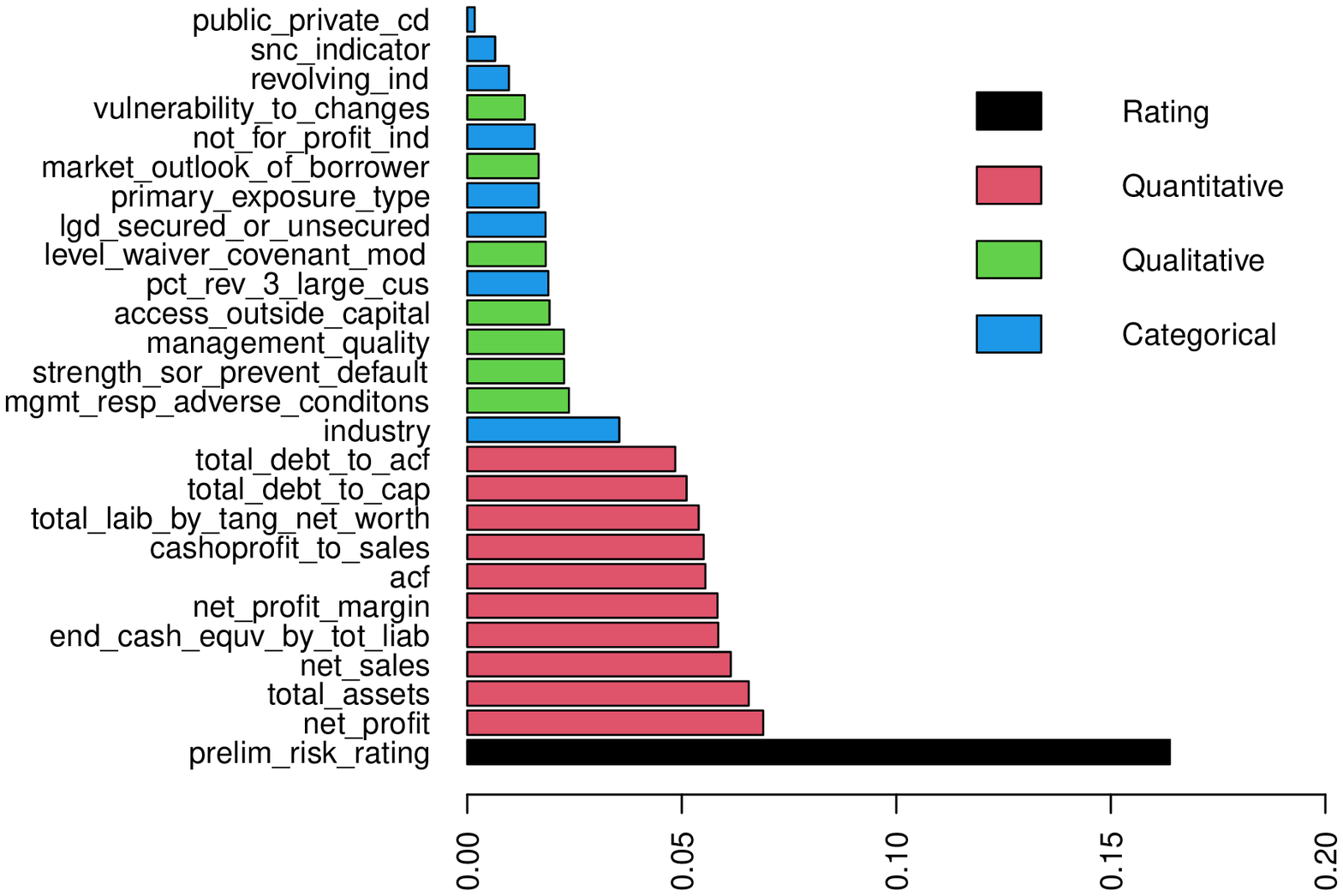}
\floatfoot{The figure shows the relative importance of all variables in the model based on the Gini index, also know as impurity measure.}
\end{figure}

\begin{figure}[htb]
\caption{Variable Importance for the Random Forest without the Scorecard in the Model}
\label{f:importance_noscore}
\includegraphics[width=1\textwidth]{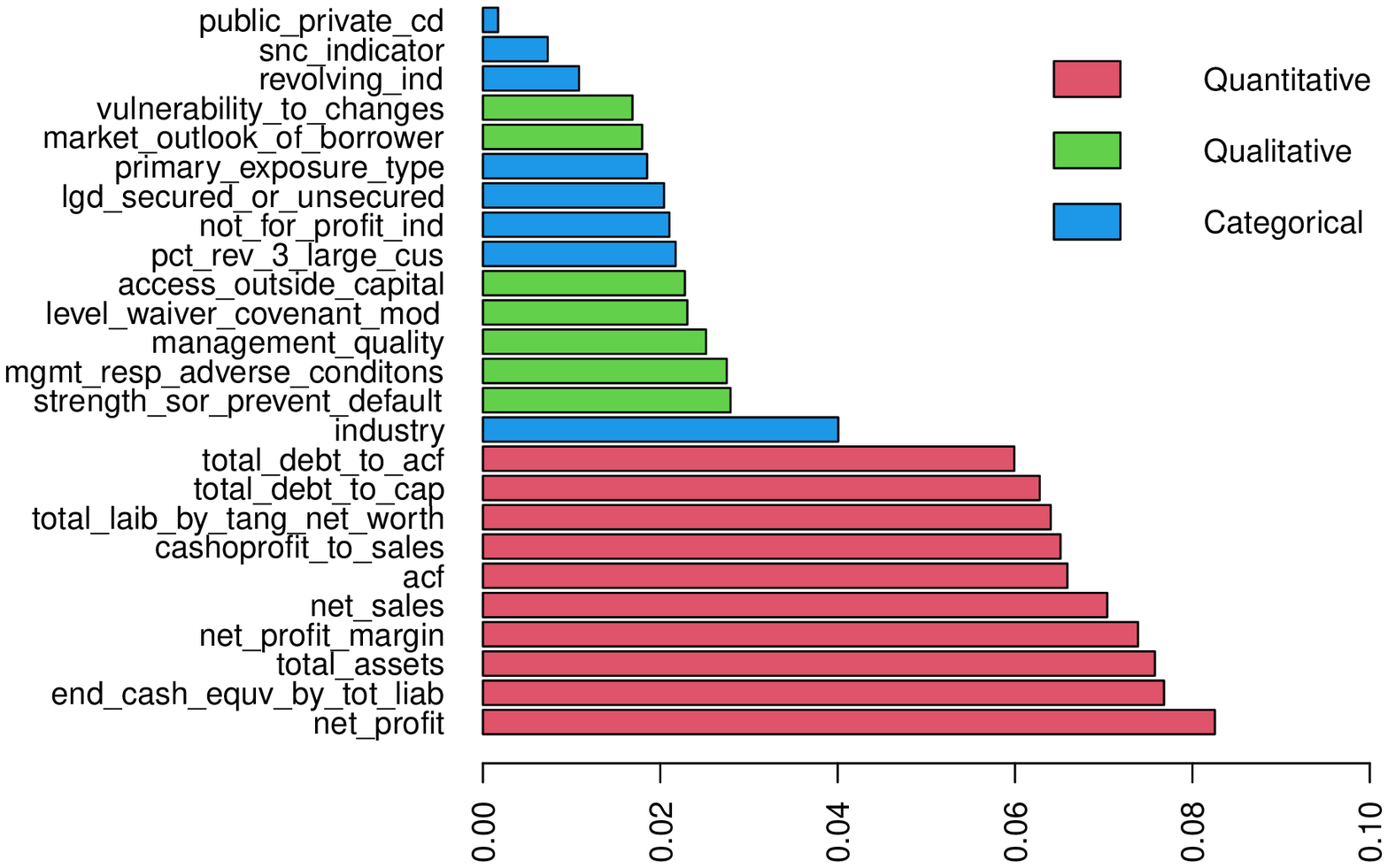}
\floatfoot{The figure shows the relative importance of all variables in the model based on the Gini index, also know as impurity measure.}
\end{figure}

Next let us consider the relative importance of the different predictors in the Random Forest model. We display the variable importance for the model with and without accounting for the scorecard rating in Figure \ref{f:importance_full} and Figure \ref{f:importance_noscore} respectively. There is no unique way to present variable importance in Random Forest models but a number of approaches are popular. The \textit{Gini Impurity Index} is a procedure described by \citet{breiman2017classification} and \citet{friedman2001elements}. Suppose we have $C$ classes and $p(i)$ is the probability of obtaining an observation with class $i=1,\dots C$. For each node $\eta$ the Gini Impurity Index is defined as:

\begin{equation}
    GI_\eta = \sum_{i = 1}^C p(i)(1-p(i))
\end{equation}

In order to obtain an importance measure from the index, first we must calculate the decrease in the impurity for each node in the tree. The second step to calculate the total decrease for each variable (using the nodes split by this variable), which is the sum of the decrease for each variable weighted by the number of samples in each node. In a Random Forest context we then take the average decrease over all trees.  This concept is related to entropy and in essence measures the likelihood of incorrectly classifying a new observation. In this context variable importance then ranks the variables according to their impact on their classification potential based on decrease in impurity (an unconditional model would rely on the empirical distribution over classes only).

\begin{center}
[Figure \ref{f:importance_full} here]
\end{center}

Figure \ref{f:importance_full} displays the variable importance for the model which includes the scorecard risk rating (prelim risk rating). Notice that it dominates all other variables in terms of predicting the final rating. This is not surprising and to some extent is simply a matter of internal consistency. As we have discussed before, notching is a relatively limited phenomenon and only in a few loans do we observe major deviations of the final rating away from the preliminary rating. Nevertheless, the other predictors contain information which helps determine the final rating. The most important block of variables is the set of financial variables followed by a mix of categorical and qualitative variables. The three most important quantitative variables are net profit, total assets and net sales. Industry is the most important categorical variable and it may indicate additional heterogeneity across industries (which is explored further in the next section). It is particularly noteworthy that after industry, qualitative variables related to management quality are also important. In particular the qualitative assessment of how well the management team is likely to respond (or did respond in the past) to adverse shocks plays an important role in assigning the final risk rating.

\begin{center}
[Figure \ref{f:importance_noscore} here]
\end{center}

Figure \ref{f:importance_noscore} displays the variable importance for the model which does not include the scorecard risk rating. As expected, the importance of each variable increases. The group level ordering of the variables is preserved but there are differences in the importance of each individual variable. Thus, the most important group of variables consists of the financial variables. They jointly increase in importance while the importance of the other variables is comparable to what it was in the previous figure. This indicates that the scorecard relies largely on financial variables and once we take the scorecard risk rating out of the model, our machine learning model replicates to some extent the information contained in the scorecard rating. The fact that the ranking of the importance of the financial variables changes however also indicates that the weights applied to the financial variables by the Random Forest may differ from that of the scorecard. There are a number of explanations for this including the fact that our machine learning algorithm is trained on this particular dataset while the scorecard weights are determined ex ante on other datasets.

\begin{figure}[htb]
\caption{Random Forest prediction Accuracy improvement by groups of variables}
\label{f:importance_increase}
\includegraphics[width=1\textwidth]{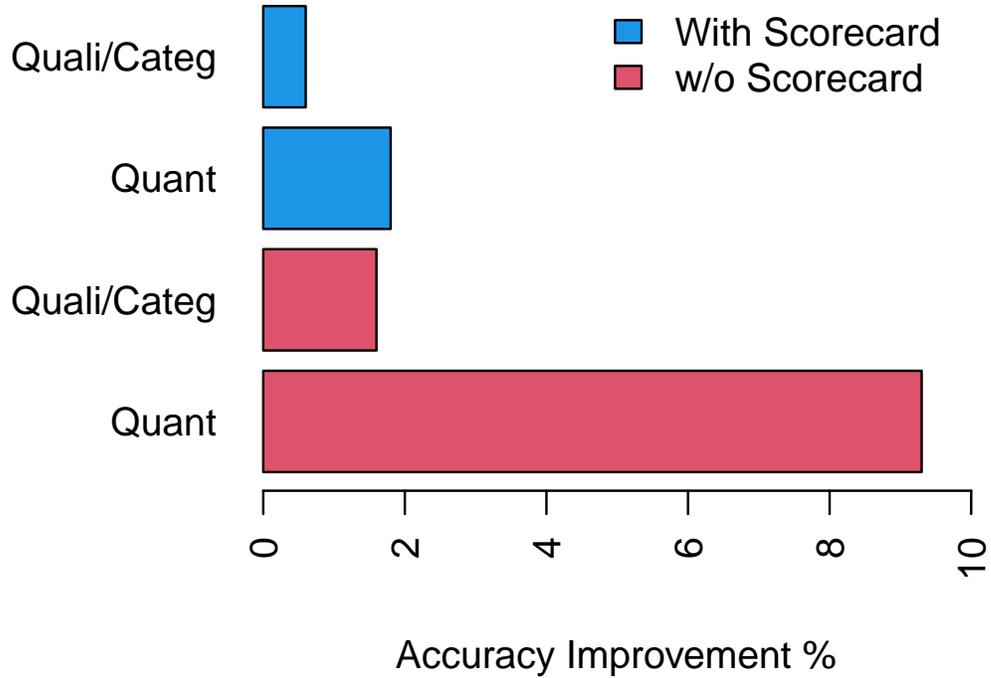}
\floatfoot{The figure shows the Random Forest 5-fold cross-validation prediction improvement of adding a group of variables to the model. For example, if we have a model only with quantitative variables and we add the qualitative/categorical variables we have an improvement as the one presented in the Quali/Quant bars. The blue bars are for the model with the Scorecard as predictor and the red bars are for the model without the Scorecard as predictor.}
\end{figure}

\begin{center}
[Figure \ref{f:importance_increase} here]
\end{center}

Figure \ref{f:importance_increase} shows the Accuracy increment of different variable groups in each model in a 5-fold cross-validation experiment. We can think of this as an experiment where we remove or add a group of variables to the model. In the model with the scorecard, the qualitative/categorical group increases the Accuracy by only 0.6\% from a model with only the quantitative variables, which increase the Accuracy by 1.8\%. The increments are bigger for the models without the Scorecard. The qualitative/categorical groups increases the Accuracy by 1.56\% and the quantitative variables increase it by more than 9.3\%. These results show that the Random Forest has a very good rearranging capacity when some variables are removed. Even if we remove the quantitative variables in the scorecard model, the result is under 2\% worse in a model only with categorical variables. This indicates that the scorecard rating acts as a summary statistic which encapsulates the quantitative information. We should nevertheless not under-emphasize the importance of the qualitative variables. In the model without the scorecard and the quantitative variables the Accuracy falls from 0.946 to 0.853, which is still a reasonable result compared to the MNL LASSO benchmark. Random forests perform relatively well even without the preliminary rating and the quantitative variables, which is a very impressive result. The relevance of qualitative and categorical variables was studied by \citet{grunert2005role} in a context of default prediction. Our results show that manager may be taking this information in to account to adjust the ratings, but a nonlinear model only loses 0.6\% (1.56\% without the preliminary rating) accuracy if we remove these variables due to is ability to explore relations that go beyond simple correlation. In other words, qualitative variables may be reflected in quantitative variables in a nonlinear way.   

\section{Heterogeneity}

In the discussion of the results above we noticed that the success of our machine learning predictions of the risk assigned by the human managers varied with the risk categories. This may indicate that there is unobserved heterogeneity that remains to be understood and potentially improved upon. Some immediate dimensions of heterogeneity are the managers themselves, the industry in which the client operates or the time at which the loan is evaluated. We shall consider each in turn.

\begin{center}
[Table \ref{tab:managers} here]
\end{center}

First, we explore the extent to which manager ``fixed effects" may be driving the results in addition to the observable characteristics of the loans. Managers may be subject to various degrees of bias such as excessive optimism or pessimism. Since each manager reviews more than one loan we can try to associate the deviations of our predictions from the rating choices of the managers . In Table \ref{tab:managers} we compute the absolute prediction error for each class and regress it against indicators for each manager. In total we have 328 managers that were involved in the rating process. Table  \ref{tab:managers} reports the number of significant coefficients at different levels of significance. Thus, at the 5\% we found 14 managers to enter the regression with significant coefficients. This shows that a small number of managers is responsible for the majority of the deviations observed. We then consider the average value of the significant coefficients and split the results by positive and negative increments. In the model with the scorecard ratings we have no negative coefficients and the average of the positive coefficients indicates that the managers who deviate significantly from the ratings assigned by the machine learning algorithm do so by about 3 notches. This indicates that for a subset of managers our model fails to accurately capture their risk rating behavior. For the model without the scorecard risk rating we also estimate negative coefficients. This can be interpreted as saying that some managers align themselves even closer to the predictions of the machine learning algorithm than the average manager. For example at the 5\% significance these managers are an average of 0.37 notches closer to the predictions of the algorithm than the average manager.

\begin{center}
[Table \ref{tab:industry} here]
\end{center}

Second, we need to consider whether the heterogeneity may depend on industry. It is perhaps the case that the riskiness of loans in some industries is more problematic to assess and thus subject to increased uncertainty that is not easy to quantify. In Table \ref{tab:industry} we compute the out of sample error from the cross-validated predictions and regress the mean absolute error against a set of industry indicators. In the data we record loans for firms categorized into 24 different industries. We report those that are significant at the usual levels of 1\%, 5\% and 10\%. We notice that we do better at predicting the final ratings in construction or retail and worse at predicting the ratings in manufacturing and finance. Recall that this is not a statement about the objective default probabilities in those sectors but rather about our ability to replicate the rating choices of the human managers evaluating loans in those sectors. It is also interesting to note that the model which includes the scorecard rating tends to perform better, but at the same time we observe more industries which have significant deviations. This might indicate a weakness of the scorecard ratings which may not accurately differentiate risk by industry type, something which the machine learning algorithms are compensating for and which was anticipated by the earlier results which showed the industry variable to be an important predictor in the Random Forest model. 

\begin{center}
[Table \ref{tab:industry} here]
\end{center}

Third, we consider whether the machine learning algorithms may have different performance on subsets of the data. One important partition is by the year in which the loan was evaluated. It is of course possible that managers evaluate the riskiness of a loan differently depending on other factors such as the macroeconomy or beliefs about future aggregate shocks. In Table \ref{tab:industry} we report the performance of the Random Forest computed for both the model with and without the scorecard assigned rating for each year in our sample. The Accuracy of our algorithm improves over time even though the sample size per year stays about the same and the model is trained for each year separately. This indicates that it is not a feature of our method, but rather a behavior change on behalf of the managers. We might conclude that over time the managers are becoming ``more algorithmic" and less discretionary. The Accuracy improves from 91.2\% in 2010 to 96.4\% in 2018 for the model which includes the scorecard rating. An Accuracy of 95.8\% is also obtained in 2018 for the model without the scorecard rating. It thus seems that the models with and without the scorecard also tend to converge over time.  We might expect that over this period the scorecard algorithm itself to have changed and perhaps become more reliable. Likewise it is possible that managers have come to rely more on the scorecard predictions which also removed the need for additional managerial discretion. Our data however only allows us to speculate on these reasons and we do not have additional evidence to draw definitive conclusions on what may have lead to this behavior change.

\section{Conclusion}

This paper explores the extent to which machine learning algorithms can replicate a part of the process of assigning risk ratings to commercial loans. First, a typical commercial bank relies on an algorithmic process and uses a scorecard to assign a preliminary risk rating to each company. Second, the companies are reviewed by a human manager in light of all available information, including the preliminary risk rating, and a final risk rating is then assigned by the human manager. In practice, the final risk rating may differ by several ``notches" from the preliminary rating. We constructed several algorithms (Multinomial Logistic LASSO, Neural Networks, Gradient Boosting, Random Forest) to predict the final risk rating assigned by the human manager. We have found that the Random Forest can predict the final risk rating with a degree of Accuracy in excess of 95\%. 

Even though financial variables are already included in the scorecard rating, they nevertheless contain additional information which is incorporated by the prediction algorithm. Qualitative variables such as assessments of managerial quality are also important in making the final determination. We also document heterogeneity in the departure of our predictions from the ratings assigned by the managers. It seems that a small number of managers is responsible for many of the differences in the predictions and that predictions may also differ by industry. Over time it appears that managers are employing less discretion which might indicate the improved performance of or reliance on algorithmic means such as the scorecard.

The results in this paper show that this particular task executed by highly skilled bank managers may in fact be easily replicated by relatively simple algorithms. The performance of these algorithms could be improved by fine tuning to account for differences across industries and of course could be easily extended to include additional goals such as incorporating considerations of fairness in lending practices or to promote other social goals.

\clearpage

\bibliographystyle{unsrtnat}
\bibliography{ref} 

\newpage

\newpage

\newpage

\newpage

\newpage

\begin{sidewaystable}[htb]
\caption{Confusion Matrix of the Managers against the Scorecard}
\label{tab:cms}
\centering
\resizebox{\textwidth}{!}{%
\begin{tabular}{cc|c|c|c|c|c|c|c|c|c|c|c|c|c|c|c|}
\multirowcell{17}{\rotatebox[origin=c]{90}{\textbf{Manager Rating}}}  &  \multicolumn{16}{c}{\textbf{Scorecard Rating}}\\
&&&&&&&&&&&&&&&&\\\hhline{~|-|-|-|-|-|-|-|-|-|-|-|-|-|-|-|-|}\noalign{\vskip.2pt}
 & \cellcolor[HTML]{E3DCCC} \cellcolor[HTML]{E3DCCC}  & \cellcolor[HTML]{E3DCCC} 2 & \cellcolor[HTML]{E3DCCC} 3 & \cellcolor[HTML]{E3DCCC} 4 & \cellcolor[HTML]{E3DCCC} 5 & \cellcolor[HTML]{E3DCCC} 6 & \cellcolor[HTML]{E3DCCC} 7 & \cellcolor[HTML]{E3DCCC} 8 & \cellcolor[HTML]{E3DCCC} 9 & \cellcolor[HTML]{E3DCCC} 10 & \cellcolor[HTML]{E3DCCC} 11 & \cellcolor[HTML]{E3DCCC} 12 & \cellcolor[HTML]{E3DCCC} 13 & \cellcolor[HTML]{E3DCCC} 14 & \cellcolor[HTML]{E3DCCC} 15 & \cellcolor[HTML]{E3DCCC} Subtotal \\ 
   \hhline{~|-|-|-|-|-|-|-|-|-|-|-|-|-|-|-|-|}\noalign{\vskip.2pt}
 & \cellcolor[HTML]{E3DCCC} 2 & \cellcolor[HTML]{0000ff}{\color[HTML]{FFFFFF} 14 } & \cellcolor[HTML]{FFFDFD} 11 & 0 & 0 & 0 & 0 & 0 & 0 & 0 & 0 & 0 & 0 & 0 & 0 & \cellcolor[HTML]{E3DCCC} 25 \\ 
   \hhline{~|-|-|-|-|-|-|-|-|-|-|-|-|-|-|-|-|}\noalign{\vskip.2pt}
 & \cellcolor[HTML]{E3DCCC} 3 & 0 & \cellcolor[HTML]{0000ff}{\color[HTML]{FFFFFF} 37 } & \cellcolor[HTML]{FFFDFD} 12 & \cellcolor[HTML]{FFFDFD} 10 & \cellcolor[HTML]{FFFEFE} 2 & 0 & 0 & 0 & 0 & 0 & 0 & 0 & 0 & 0 & \cellcolor[HTML]{E3DCCC} 61 \\ 
  \hhline{~|-|-|-|-|-|-|-|-|-|-|-|-|-|-|-|-|}\noalign{\vskip.2pt} 
 & \cellcolor[HTML]{E3DCCC} 4 & \cellcolor[HTML]{FFFAFA} 31 & \cellcolor[HTML]{FFF6F6} 53 & \cellcolor[HTML]{0000ff}{\color[HTML]{FFFFFF} 163 } & \cellcolor[HTML]{FFFFFF} 1 & \cellcolor[HTML]{FFF5F5} 60 & \cellcolor[HTML]{FFFEFE} 3 & \cellcolor[HTML]{FFFCFC} 17 & \cellcolor[HTML]{FFFEFE} 2 & \cellcolor[HTML]{FFF9F9} 36 & 0 & \cellcolor[HTML]{FFFFFF} 1 & 0 & 0 & 0 & \cellcolor[HTML]{E3DCCC} 367 \\ 
   \hhline{~|-|-|-|-|-|-|-|-|-|-|-|-|-|-|-|-|}\noalign{\vskip.2pt}
 & \cellcolor[HTML]{E3DCCC} 5 & 0 & \cellcolor[HTML]{FFFDFD} 13 & \cellcolor[HTML]{FFEEEE} 102 & \cellcolor[HTML]{0000ff}{\color[HTML]{FFFFFF} 131 } & \cellcolor[HTML]{FFFDFD} 8 & \cellcolor[HTML]{FFFDFD} 12 & 0 & \cellcolor[HTML]{FFFEFE} 3 & \cellcolor[HTML]{FFFEFE} 3 & 0 & 0 & \cellcolor[HTML]{FFFDFD} 9 & \cellcolor[HTML]{FFFAFA} 29 & \cellcolor[HTML]{FFFEFE} 4 & \cellcolor[HTML]{E3DCCC} 314 \\ 
   \hhline{~|-|-|-|-|-|-|-|-|-|-|-|-|-|-|-|-|}\noalign{\vskip.2pt}
 & \cellcolor[HTML]{E3DCCC} 6 & 0 & \cellcolor[HTML]{FFFAFA} 30 & \cellcolor[HTML]{FFEEEE} 100 & \cellcolor[HTML]{FFF2F2} 81 & \cellcolor[HTML]{0000ff}{\color[HTML]{FFFFFF} 461 } & \cellcolor[HTML]{FFFAFA} 31 & \cellcolor[HTML]{FFF2F2} 79 & \cellcolor[HTML]{FFF1F1} 87 & \cellcolor[HTML]{FFFEFE} 4 & \cellcolor[HTML]{FFFBFB} 20 & \cellcolor[HTML]{FFF8F8} 41 & \cellcolor[HTML]{FFF9F9} 35 & \cellcolor[HTML]{FFFEFE} 2 & \cellcolor[HTML]{FFF4F4} 63 & \cellcolor[HTML]{E3DCCC} 1034 \\ 
  \hhline{~|-|-|-|-|-|-|-|-|-|-|-|-|-|-|-|-|}\noalign{\vskip.2pt} 
 & \cellcolor[HTML]{E3DCCC} 7 & 0 & 0 & \cellcolor[HTML]{FFFAFA} 30 & \cellcolor[HTML]{FFEAEA} 128 & \cellcolor[HTML]{FFF3F3} 72 & \cellcolor[HTML]{0000ff}{\color[HTML]{FFFFFF} 1720 } & \cellcolor[HTML]{FFF9F9} 36 & \cellcolor[HTML]{FF9D9D} 604 & \cellcolor[HTML]{FFF3F3} 69 & \cellcolor[HTML]{FFF6F6} 56 & \cellcolor[HTML]{FFFCFC} 15 & \cellcolor[HTML]{FFFCFC} 15 & \cellcolor[HTML]{FFF9F9} 36 & \cellcolor[HTML]{FFFDFD} 11 & \cellcolor[HTML]{E3DCCC} 2792 \\ 
   \hhline{~|-|-|-|-|-|-|-|-|-|-|-|-|-|-|-|-|}\noalign{\vskip.2pt}
 & \cellcolor[HTML]{E3DCCC} 8 & 0 & 0 & 0 & \cellcolor[HTML]{FFF9F9} 36 & \cellcolor[HTML]{FFF5F5} 60 & \cellcolor[HTML]{FFBEBE} 399 & \cellcolor[HTML]{0000ff}{\color[HTML]{FFFFFF} 991 } & \cellcolor[HTML]{FF8D8D} 700 & \cellcolor[HTML]{FFEAEA} 125 & \cellcolor[HTML]{FFF0F0} 92 & \cellcolor[HTML]{FFF7F7} 45 & \cellcolor[HTML]{FFFDFD} 12 & \cellcolor[HTML]{FFF6F6} 53 & \cellcolor[HTML]{FFF3F3} 74 & \cellcolor[HTML]{E3DCCC} 2587 \\ 
  \hhline{~|-|-|-|-|-|-|-|-|-|-|-|-|-|-|-|-|}\noalign{\vskip.2pt} 
 & \cellcolor[HTML]{E3DCCC} 9 & 0 & 0 & 0 & 0 & \cellcolor[HTML]{FFFBFB} 21 & \cellcolor[HTML]{FFD7D7} 245 & \cellcolor[HTML]{FFCFCF} 295 & \cellcolor[HTML]{0000ff}{\color[HTML]{FFFFFF} 7671 } & \cellcolor[HTML]{FFCFCF} 294 & \cellcolor[HTML]{FFE3E3} 171 & \cellcolor[HTML]{FFF3F3} 69 & \cellcolor[HTML]{FFFBFB} 20 & \cellcolor[HTML]{FFF4F4} 67 & \cellcolor[HTML]{FFF2F2} 79 & \cellcolor[HTML]{E3DCCC} 8932 \\ 
   \hhline{~|-|-|-|-|-|-|-|-|-|-|-|-|-|-|-|-|}\noalign{\vskip.2pt}
 & \cellcolor[HTML]{E3DCCC} 10 & 0 & 0 & 0 & 0 & \cellcolor[HTML]{FFFFFF} 1 & \cellcolor[HTML]{FFFCFC} 17 & \cellcolor[HTML]{FFF4F4} 64 & \cellcolor[HTML]{FF0000} 1573 & \cellcolor[HTML]{0000ff}{\color[HTML]{FFFFFF} 2446 } & \cellcolor[HTML]{FFD9D9} 233 & \cellcolor[HTML]{FFE4E4} 162 & \cellcolor[HTML]{FFE6E6} 155 & \cellcolor[HTML]{FFEAEA} 127 & \cellcolor[HTML]{FFD7D7} 242 & \cellcolor[HTML]{E3DCCC} 5020 \\ 
   \hhline{~|-|-|-|-|-|-|-|-|-|-|-|-|-|-|-|-|}\noalign{\vskip.2pt}
 & \cellcolor[HTML]{E3DCCC} 11 & 0 & 0 & 0 & 0 & 0 & \cellcolor[HTML]{FFFEFE} 3 & \cellcolor[HTML]{FFF8F8} 44 & \cellcolor[HTML]{FF0202} 1560 & \cellcolor[HTML]{FF8686} 741 & \cellcolor[HTML]{0000ff}{\color[HTML]{FFFFFF} 2535 } & \cellcolor[HTML]{FFE5E5} 160 & \cellcolor[HTML]{FFE0E0} 192 & \cellcolor[HTML]{FFEBEB} 121 & \cellcolor[HTML]{FFDADA} 227 & \cellcolor[HTML]{E3DCCC} 5583 \\ 
  \hhline{~|-|-|-|-|-|-|-|-|-|-|-|-|-|-|-|-|}\noalign{\vskip.2pt} 
 & \cellcolor[HTML]{E3DCCC} 12 & 0 & 0 & 0 & 0 & 0 & \cellcolor[HTML]{FFF2F2} 81 & \cellcolor[HTML]{FFFCFC} 16 & \cellcolor[HTML]{FF8F8F} 691 & \cellcolor[HTML]{FF8585} 753 & \cellcolor[HTML]{FF4242} 1166 & \cellcolor[HTML]{0000ff}{\color[HTML]{FFFFFF} 2037 } & \cellcolor[HTML]{FFB7B7} 440 & \cellcolor[HTML]{FFC6C6} 348 & \cellcolor[HTML]{FFBEBE} 400 & \cellcolor[HTML]{E3DCCC} 5932 \\ 
   \hhline{~|-|-|-|-|-|-|-|-|-|-|-|-|-|-|-|-|}\noalign{\vskip.2pt}
 & \cellcolor[HTML]{E3DCCC} 13 & 0 & 0 & 0 & 0 & 0 & \cellcolor[HTML]{FFFEFE} 4 & 0 & \cellcolor[HTML]{FFF0F0} 88 & \cellcolor[HTML]{FFF6F6} 54 & \cellcolor[HTML]{FFD8D8} 239 & \cellcolor[HTML]{FFCCCC} 311 & \cellcolor[HTML]{0000ff}{\color[HTML]{FFFFFF} 879 } & \cellcolor[HTML]{FFF9F9} 35 & \cellcolor[HTML]{FFEAEA} 129 & \cellcolor[HTML]{E3DCCC} 1739 \\ 
   \hhline{~|-|-|-|-|-|-|-|-|-|-|-|-|-|-|-|-|}\noalign{\vskip.2pt}
 & \cellcolor[HTML]{E3DCCC} 14 & 0 & 0 & 0 & 0 & 0 & 0 & 0 & \cellcolor[HTML]{FFFFFF} 1 & \cellcolor[HTML]{FFF7F7} 47 & \cellcolor[HTML]{FFF4F4} 64 & \cellcolor[HTML]{FFEEEE} 104 & \cellcolor[HTML]{FFF1F1} 83 & \cellcolor[HTML]{0000ff}{\color[HTML]{FFFFFF} 484 } & \cellcolor[HTML]{FFE7E7} 148 & \cellcolor[HTML]{E3DCCC} 931 \\ 
   \hhline{~|-|-|-|-|-|-|-|-|-|-|-|-|-|-|-|-|}\noalign{\vskip.2pt}
 & \cellcolor[HTML]{E3DCCC} 15 & 0 & 0 & 0 & 0 & 0 & 0 & 0 & \cellcolor[HTML]{FFE3E3} 169 & \cellcolor[HTML]{FFF7F7} 48 & \cellcolor[HTML]{FFE3E3} 169 & \cellcolor[HTML]{FFF3F3} 71 & \cellcolor[HTML]{FFC5C5} 354 & \cellcolor[HTML]{FFC4C4} 364 & \cellcolor[HTML]{0000ff}{\color[HTML]{FFFFFF} 957 } & \cellcolor[HTML]{E3DCCC} 2132 \\ 
  \hhline{~|-|-|-|-|-|-|-|-|-|-|-|-|-|-|-|-|}\noalign{\vskip.2pt} 
 & \cellcolor[HTML]{E3DCCC} Subtotal & \cellcolor[HTML]{E3DCCC} 45 & \cellcolor[HTML]{E3DCCC} 144 & \cellcolor[HTML]{E3DCCC} 407 & \cellcolor[HTML]{E3DCCC} 387 & \cellcolor[HTML]{E3DCCC} 685 & \cellcolor[HTML]{E3DCCC} 2515 & \cellcolor[HTML]{E3DCCC} 1542 & \cellcolor[HTML]{E3DCCC} 13149 & \cellcolor[HTML]{E3DCCC} 4620 & \cellcolor[HTML]{E3DCCC} 4745 & \cellcolor[HTML]{E3DCCC} 3016 & \cellcolor[HTML]{E3DCCC} 2194 & \cellcolor[HTML]{E3DCCC} 1666 & \cellcolor[HTML]{E3DCCC} 2334 & \cellcolor[HTML]{E3DCCC} 37449 \\ 
   \hhline{~|-|-|-|-|-|-|-|-|-|-|-|-|-|-|-|-|}\noalign{\vskip.2pt}
\end{tabular}
}
\end{sidewaystable}

\newpage

\begin{table}[ht]
\caption{List of Variables}
\label{tab:variables}
\centering
\begin{tabular}{lc}
  \hline
Name & Type \\ 
  \hline
primary\_exposure\_type & Categorical \\ 
  pct\_rev\_3\_large\_cus & Categorical \\ 
  public\_private\_cd & Categorical \\ 
  snc\_indicator & Categorical \\ 
  not\_for\_profit\_ind & Categorical \\ 
  revolving\_ind & Categorical \\ 
  lgd\_secured\_or\_unsecured & Categorical \\ 
  industry & Categorical \\ 
  access\_outside\_capital & Qualitative \\ 
  level\_waiver\_covenant\_mod & Qualitative \\ 
  management\_quality & Qualitative \\ 
  market\_outlook\_of\_borrower & Qualitative \\ 
  mgmt\_resp\_adverse\_conditons & Qualitative \\ 
  strength\_sor\_prevent\_default & Qualitative \\ 
  vulnerability\_to\_changes & Qualitative \\ 
  net\_sales & Quantitative \\ 
  net\_profit\_margin & Quantitative \\ 
  cashoprofit\_to\_sales & Quantitative \\ 
  total\_assets & Quantitative \\ 
  total\_laib\_by\_tang\_net\_worth & Quantitative \\ 
  acf & Quantitative \\ 
  total\_debt\_to\_cap & Quantitative \\ 
  total\_debt\_to\_acf & Quantitative \\ 
  end\_cash\_equv\_by\_tot\_liab & Quantitative \\ 
  net\_profit & Quantitative \\ 
  prelim\_risk\_rating & Rating \\ 
  final\_risk\_rating & Rating \\ 
   \hline
\end{tabular}
\end{table}

\newpage

\begin{table}
\begin{threeparttable}
\caption{Cross Validation Prediction Results }
\label{tab:cv}
\resizebox{\textwidth}{!}{%
\begin{tabular}{lccccc}
\hline
          & \multicolumn{5}{c}{Results with the scorecard rating as predictor}    \\
          & RMSE    & Accuracy  & AccuracyLower  & AccuracyUpper & AccuracyPValue \\ \hline
MNL LASSO & 1.361   & 0.537     & 0.532          & 0.542         & 0.000          \\
NN        & 0.447   & 0.934     & 0.931          & 0.936         & 0.000          \\
Boosting  & 0.332   & 0.958     & 0.955          & 0.960         & 0.000          \\
R. Forest & 0.326   & 0.959     & 0.957          & 0.961         & 0.000          \\ \hline
          &         &           &                &               &                \\
          & \multicolumn{5}{c}{Results without the scorecard rating as predictor} \\ \hline
MNL Lasso & 1.561   & 0.473     & 0.468          & 0.479         & 0.000          \\
NN        & 0.514   & 0.905     & 0.902          & 0.908         & 0.000          \\
Boosting  & 0.371   & 0.943     & 0.941          & 0.946         & 0.000          \\
R. Forest & 0.365   & 0.946     & 0.943          & 0.948         & 0.000          \\ \hline
\end{tabular}%
}%
\begin{tablenotes}
\small
\item The table shows the prediction accuracy measured on a 5-fold cross-validation for the Multinomial Logistic LASSO, Neural Network, Gradient Boosting and Random Forest. 
\end{tablenotes}
\end{threeparttable}
\end{table}

\newpage

\begin{sidewaystable}[htb]
\caption{Confusion Matrix of the Random Forest with the Scorecard as Predictor}
\label{tab:cmrf_full}
\centering
\resizebox{\textwidth}{!}{%
\begin{tabular}{cc|c|c|c|c|c|c|c|c|c|c|c|c|c|c|c|}
\multirowcell{17}{\rotatebox[origin=c]{90}{\textbf{Model Rating}}}  &  \multicolumn{16}{c}{\textbf{Manager Rating}}\\
&&&&&&&&&&&&&&&&\\\hhline{~|-|-|-|-|-|-|-|-|-|-|-|-|-|-|-|-|}\noalign{\vskip.2pt}
 & \cellcolor[HTML]{E3DCCC} \cellcolor[HTML]{E3DCCC}  & \cellcolor[HTML]{E3DCCC} 2 & \cellcolor[HTML]{E3DCCC} 3 & \cellcolor[HTML]{E3DCCC} 4 & \cellcolor[HTML]{E3DCCC} 5 & \cellcolor[HTML]{E3DCCC} 6 & \cellcolor[HTML]{E3DCCC} 7 & \cellcolor[HTML]{E3DCCC} 8 & \cellcolor[HTML]{E3DCCC} 9 & \cellcolor[HTML]{E3DCCC} 10 & \cellcolor[HTML]{E3DCCC} 11 & \cellcolor[HTML]{E3DCCC} 12 & \cellcolor[HTML]{E3DCCC} 13 & \cellcolor[HTML]{E3DCCC} 14 & \cellcolor[HTML]{E3DCCC} 15 & \cellcolor[HTML]{E3DCCC} Subtotal \\ 
   \hhline{~|-|-|-|-|-|-|-|-|-|-|-|-|-|-|-|-|}\noalign{\vskip.2pt}
 & \cellcolor[HTML]{E3DCCC} 2 & \cellcolor[HTML]{0000ff}{\color[HTML]{FFFFFF} 25 } & 0 & 0 & 0 & 0 & 0 & 0 & 0 & 0 & 0 & 0 & 0 & 0 & 0 & \cellcolor[HTML]{E3DCCC} 25 \\ 
   \hhline{~|-|-|-|-|-|-|-|-|-|-|-|-|-|-|-|-|}\noalign{\vskip.2pt}
 & \cellcolor[HTML]{E3DCCC} 3 & 0 & \cellcolor[HTML]{0000ff}{\color[HTML]{FFFFFF} 56 } & \cellcolor[HTML]{FFFBFB} 3 & 0 & 0 & 0 & 0 & 0 & 0 & 0 & 0 & 0 & 0 & 0 & \cellcolor[HTML]{E3DCCC} 59 \\ 
   \hhline{~|-|-|-|-|-|-|-|-|-|-|-|-|-|-|-|-|}\noalign{\vskip.2pt}
 & \cellcolor[HTML]{E3DCCC} 4 & 0 & \cellcolor[HTML]{FFFAFA} 4 & \cellcolor[HTML]{0000ff}{\color[HTML]{FFFFFF} 347 } & \cellcolor[HTML]{FFF1F1} 9 & \cellcolor[HTML]{FFFFFF} 1 & \cellcolor[HTML]{FFFFFF} 1 & 0 & 0 & 0 & 0 & 0 & 0 & 0 & 0 & \cellcolor[HTML]{E3DCCC} 362 \\ 
   \hhline{~|-|-|-|-|-|-|-|-|-|-|-|-|-|-|-|-|}\noalign{\vskip.2pt}
 & \cellcolor[HTML]{E3DCCC} 5 & 0 & 0 & \cellcolor[HTML]{FFF5F5} 7 & \cellcolor[HTML]{0000ff}{\color[HTML]{FFFFFF} 290 } & \cellcolor[HTML]{FFFBFB} 3 & \cellcolor[HTML]{FFFFFF} 1 & 0 & 0 & 0 & 0 & 0 & 0 & 0 & 0 & \cellcolor[HTML]{E3DCCC} 301 \\ 
   \hhline{~|-|-|-|-|-|-|-|-|-|-|-|-|-|-|-|-|}\noalign{\vskip.2pt}
 & \cellcolor[HTML]{E3DCCC} 6 & 0 & 0 & \cellcolor[HTML]{FFFBFB} 3 & \cellcolor[HTML]{FFF1F1} 9 & \cellcolor[HTML]{0000ff}{\color[HTML]{FFFFFF} 1009 } & \cellcolor[HTML]{FFD5D5} 26 & \cellcolor[HTML]{FFF8F8} 5 & 0 & 0 & 0 & 0 & 0 & 0 & 0 & \cellcolor[HTML]{E3DCCC} 1052 \\ 
   \hhline{~|-|-|-|-|-|-|-|-|-|-|-|-|-|-|-|-|}\noalign{\vskip.2pt}
 & \cellcolor[HTML]{E3DCCC} 7 & 0 & \cellcolor[HTML]{FFFFFF} 1 & \cellcolor[HTML]{FFFAFA} 4 & \cellcolor[HTML]{FFFBFB} 3 & \cellcolor[HTML]{FFE7E7} 15 & \cellcolor[HTML]{0000ff}{\color[HTML]{FFFFFF} 2650 } & \cellcolor[HTML]{FFACAC} 51 & \cellcolor[HTML]{FFC6C6} 35 & \cellcolor[HTML]{FFFAFA} 4 & \cellcolor[HTML]{FFFBFB} 3 & 0 & 0 & 0 & 0 & \cellcolor[HTML]{E3DCCC} 2766 \\ 
   \hhline{~|-|-|-|-|-|-|-|-|-|-|-|-|-|-|-|-|}\noalign{\vskip.2pt}
 & \cellcolor[HTML]{E3DCCC} 8 & 0 & 0 & \cellcolor[HTML]{FFFFFF} 1 & \cellcolor[HTML]{FFFDFD} 2 & \cellcolor[HTML]{FFFAFA} 4 & \cellcolor[HTML]{FFDCDC} 22 & \cellcolor[HTML]{0000ff}{\color[HTML]{FFFFFF} 2390 } & \cellcolor[HTML]{FF9090} 68 & \cellcolor[HTML]{FFE6E6} 16 & \cellcolor[HTML]{FFFDFD} 2 & 0 & \cellcolor[HTML]{FFFFFF} 1 & 0 & 0 & \cellcolor[HTML]{E3DCCC} 2506 \\ 
   \hhline{~|-|-|-|-|-|-|-|-|-|-|-|-|-|-|-|-|}\noalign{\vskip.2pt}
 & \cellcolor[HTML]{E3DCCC} 9 & 0 & 0 & \cellcolor[HTML]{FFFFFF} 1 & \cellcolor[HTML]{FFFFFF} 1 & \cellcolor[HTML]{FFFFFF} 1 & \cellcolor[HTML]{FF7C7C} 80 & \cellcolor[HTML]{FF4F4F} 107 & \cellcolor[HTML]{0000ff}{\color[HTML]{FFFFFF} 8715 } & \cellcolor[HTML]{FF0000} 155 & \cellcolor[HTML]{FF7A7A} 81 & \cellcolor[HTML]{FFCDCD} 31 & \cellcolor[HTML]{FFFDFD} 2 & 0 & 0 & \cellcolor[HTML]{E3DCCC} 9174 \\ 
   \hhline{~|-|-|-|-|-|-|-|-|-|-|-|-|-|-|-|-|}\noalign{\vskip.2pt}
 & \cellcolor[HTML]{E3DCCC} 10 & 0 & 0 & \cellcolor[HTML]{FFFFFF} 1 & 0 & 0 & \cellcolor[HTML]{FFF8F8} 5 & \cellcolor[HTML]{FFD8D8} 24 & \cellcolor[HTML]{FF8686} 74 & \cellcolor[HTML]{0000ff}{\color[HTML]{FFFFFF} 4740 } & \cellcolor[HTML]{FF3939} 120 & \cellcolor[HTML]{FFB4B4} 46 & \cellcolor[HTML]{FFFDFD} 2 & 0 & 0 & \cellcolor[HTML]{E3DCCC} 5012 \\ 
   \hhline{~|-|-|-|-|-|-|-|-|-|-|-|-|-|-|-|-|}\noalign{\vskip.2pt}
 & \cellcolor[HTML]{E3DCCC} 11 & 0 & 0 & 0 & 0 & \cellcolor[HTML]{FFFFFF} 1 & \cellcolor[HTML]{FFFDFD} 2 & \cellcolor[HTML]{FFF5F5} 7 & \cellcolor[HTML]{FFDCDC} 22 & \cellcolor[HTML]{FF9090} 68 & \cellcolor[HTML]{0000ff}{\color[HTML]{FFFFFF} 5267 } & \cellcolor[HTML]{FF7F7F} 78 & \cellcolor[HTML]{FFD0D0} 29 & \cellcolor[HTML]{FFFDFD} 2 & 0 & \cellcolor[HTML]{E3DCCC} 5476 \\ 
   \hhline{~|-|-|-|-|-|-|-|-|-|-|-|-|-|-|-|-|}\noalign{\vskip.2pt}
 & \cellcolor[HTML]{E3DCCC} 12 & 0 & 0 & 0 & 0 & 0 & \cellcolor[HTML]{FFFAFA} 4 & \cellcolor[HTML]{FFFBFB} 3 & \cellcolor[HTML]{FFE6E6} 16 & \cellcolor[HTML]{FFC5C5} 36 & \cellcolor[HTML]{FF6D6D} 89 & \cellcolor[HTML]{0000ff}{\color[HTML]{FFFFFF} 5762 } & \cellcolor[HTML]{FFBEBE} 40 & \cellcolor[HTML]{FFF5F5} 7 & \cellcolor[HTML]{FFF3F3} 8 & \cellcolor[HTML]{E3DCCC} 5965 \\ 
   \hhline{~|-|-|-|-|-|-|-|-|-|-|-|-|-|-|-|-|}\noalign{\vskip.2pt}
 & \cellcolor[HTML]{E3DCCC} 13 & 0 & 0 & 0 & 0 & 0 & \cellcolor[HTML]{FFFFFF} 1 & 0 & \cellcolor[HTML]{FFFDFD} 2 & 0 & \cellcolor[HTML]{FFDFDF} 20 & \cellcolor[HTML]{FFE9E9} 14 & \cellcolor[HTML]{0000ff}{\color[HTML]{FFFFFF} 1651 } & \cellcolor[HTML]{FFF0F0} 10 & \cellcolor[HTML]{FFF6F6} 6 & \cellcolor[HTML]{E3DCCC} 1704 \\ 
   \hhline{~|-|-|-|-|-|-|-|-|-|-|-|-|-|-|-|-|}\noalign{\vskip.2pt}
 & \cellcolor[HTML]{E3DCCC} 14 & 0 & 0 & 0 & 0 & 0 & 0 & 0 & 0 & 0 & \cellcolor[HTML]{FFFFFF} 1 & \cellcolor[HTML]{FFFFFF} 1 & \cellcolor[HTML]{FFEBEB} 13 & \cellcolor[HTML]{0000ff}{\color[HTML]{FFFFFF} 899 } & \cellcolor[HTML]{FFF8F8} 5 & \cellcolor[HTML]{E3DCCC} 919 \\ 
   \hhline{~|-|-|-|-|-|-|-|-|-|-|-|-|-|-|-|-|}\noalign{\vskip.2pt}
 & \cellcolor[HTML]{E3DCCC} 15 & 0 & 0 & 0 & 0 & 0 & 0 & 0 & 0 & \cellcolor[HTML]{FFFFFF} 1 & 0 & 0 & \cellcolor[HTML]{FFFFFF} 1 & \cellcolor[HTML]{FFEBEB} 13 & \cellcolor[HTML]{0000ff}{\color[HTML]{FFFFFF} 2113 } & \cellcolor[HTML]{E3DCCC} 2128 \\ 
   \hhline{~|-|-|-|-|-|-|-|-|-|-|-|-|-|-|-|-|}\noalign{\vskip.2pt}
 & \cellcolor[HTML]{E3DCCC} Subtotal & \cellcolor[HTML]{E3DCCC} 25 & \cellcolor[HTML]{E3DCCC} 61 & \cellcolor[HTML]{E3DCCC} 367 & \cellcolor[HTML]{E3DCCC} 314 & \cellcolor[HTML]{E3DCCC} 1034 & \cellcolor[HTML]{E3DCCC} 2792 & \cellcolor[HTML]{E3DCCC} 2587 & \cellcolor[HTML]{E3DCCC} 8932 & \cellcolor[HTML]{E3DCCC} 5020 & \cellcolor[HTML]{E3DCCC} 5583 & \cellcolor[HTML]{E3DCCC} 5932 & \cellcolor[HTML]{E3DCCC} 1739 & \cellcolor[HTML]{E3DCCC} 931 & \cellcolor[HTML]{E3DCCC} 2132 & \cellcolor[HTML]{E3DCCC} 37449 \\ 
   \hhline{~|-|-|-|-|-|-|-|-|-|-|-|-|-|-|-|-|}\noalign{\vskip.2pt}
\end{tabular}
}
\end{sidewaystable}

\newpage

\begin{sidewaystable}[htb]
\caption{Confusion Matrix of the Random Forest without the Scorecard as Predictor}
\label{tab:cmrf_noscore}
\centering
\resizebox{\textwidth}{!}{%
\begin{tabular}{cc|c|c|c|c|c|c|c|c|c|c|c|c|c|c|c|}
\multirowcell{17}{\rotatebox[origin=c]{90}{\textbf{Model Rating}}}  &  \multicolumn{16}{c}{\textbf{Manager Rating}}\\
&&&&&&&&&&&&&&&&\\\hhline{~|-|-|-|-|-|-|-|-|-|-|-|-|-|-|-|-|}\noalign{\vskip.2pt}
 & \cellcolor[HTML]{E3DCCC} \cellcolor[HTML]{E3DCCC}  & \cellcolor[HTML]{E3DCCC} 2 & \cellcolor[HTML]{E3DCCC} 3 & \cellcolor[HTML]{E3DCCC} 4 & \cellcolor[HTML]{E3DCCC} 5 & \cellcolor[HTML]{E3DCCC} 6 & \cellcolor[HTML]{E3DCCC} 7 & \cellcolor[HTML]{E3DCCC} 8 & \cellcolor[HTML]{E3DCCC} 9 & \cellcolor[HTML]{E3DCCC} 10 & \cellcolor[HTML]{E3DCCC} 11 & \cellcolor[HTML]{E3DCCC} 12 & \cellcolor[HTML]{E3DCCC} 13 & \cellcolor[HTML]{E3DCCC} 14 & \cellcolor[HTML]{E3DCCC} 15 & \cellcolor[HTML]{E3DCCC} Subtotal \\ 
   \hhline{~|-|-|-|-|-|-|-|-|-|-|-|-|-|-|-|-|}\noalign{\vskip.2pt}
 & \cellcolor[HTML]{E3DCCC} 2 & \cellcolor[HTML]{0000ff}{\color[HTML]{FFFFFF} 25 } & 0 & 0 & 0 & 0 & 0 & 0 & 0 & 0 & 0 & 0 & 0 & 0 & 0 & \cellcolor[HTML]{E3DCCC} 25 \\ 
   \hhline{~|-|-|-|-|-|-|-|-|-|-|-|-|-|-|-|-|}\noalign{\vskip.2pt}
 & \cellcolor[HTML]{E3DCCC} 3 & 0 & \cellcolor[HTML]{0000ff}{\color[HTML]{FFFFFF} 55 } & \cellcolor[HTML]{FFF8F8} 6 & 0 & 0 & 0 & 0 & 0 & 0 & 0 & 0 & 0 & 0 & 0 & \cellcolor[HTML]{E3DCCC} 61 \\ 
   \hhline{~|-|-|-|-|-|-|-|-|-|-|-|-|-|-|-|-|}\noalign{\vskip.2pt}
 & \cellcolor[HTML]{E3DCCC} 4 & 0 & \cellcolor[HTML]{FFF9F9} 5 & \cellcolor[HTML]{0000ff}{\color[HTML]{FFFFFF} 343 } & \cellcolor[HTML]{FFEFEF} 13 & \cellcolor[HTML]{FFFFFF} 1 & \cellcolor[HTML]{FFFFFF} 1 & 0 & 0 & 0 & 0 & 0 & 0 & 0 & 0 & \cellcolor[HTML]{E3DCCC} 363 \\ 
   \hhline{~|-|-|-|-|-|-|-|-|-|-|-|-|-|-|-|-|}\noalign{\vskip.2pt}
 & \cellcolor[HTML]{E3DCCC} 5 & 0 & 0 & \cellcolor[HTML]{FFF4F4} 9 & \cellcolor[HTML]{0000ff}{\color[HTML]{FFFFFF} 284 } & \cellcolor[HTML]{FFF4F4} 9 & \cellcolor[HTML]{FFFCFC} 3 & 0 & 0 & 0 & 0 & 0 & 0 & 0 & 0 & \cellcolor[HTML]{E3DCCC} 305 \\ 
   \hhline{~|-|-|-|-|-|-|-|-|-|-|-|-|-|-|-|-|}\noalign{\vskip.2pt}
 & \cellcolor[HTML]{E3DCCC} 6 & 0 & 0 & \cellcolor[HTML]{FFFDFD} 2 & \cellcolor[HTML]{FFF0F0} 12 & \cellcolor[HTML]{0000ff}{\color[HTML]{FFFFFF} 998 } & \cellcolor[HTML]{FFD6D6} 33 & \cellcolor[HTML]{FFF7F7} 7 & \cellcolor[HTML]{FFFFFF} 1 & 0 & 0 & 0 & 0 & 0 & 0 & \cellcolor[HTML]{E3DCCC} 1053 \\ 
   \hhline{~|-|-|-|-|-|-|-|-|-|-|-|-|-|-|-|-|}\noalign{\vskip.2pt}
 & \cellcolor[HTML]{E3DCCC} 7 & 0 & \cellcolor[HTML]{FFFFFF} 1 & \cellcolor[HTML]{FFF8F8} 6 & \cellcolor[HTML]{FFFCFC} 3 & \cellcolor[HTML]{FFEEEE} 14 & \cellcolor[HTML]{0000ff}{\color[HTML]{FFFFFF} 2581 } & \cellcolor[HTML]{FFC5C5} 46 & \cellcolor[HTML]{FFBFBF} 51 & \cellcolor[HTML]{FFFDFD} 2 & \cellcolor[HTML]{FFFCFC} 3 & \cellcolor[HTML]{FFFFFF} 1 & 0 & 0 & 0 & \cellcolor[HTML]{E3DCCC} 2708 \\ 
   \hhline{~|-|-|-|-|-|-|-|-|-|-|-|-|-|-|-|-|}\noalign{\vskip.2pt}
 & \cellcolor[HTML]{E3DCCC} 8 & 0 & 0 & 0 & \cellcolor[HTML]{FFFFFF} 1 & \cellcolor[HTML]{FFF4F4} 9 & \cellcolor[HTML]{FFD3D3} 35 & \cellcolor[HTML]{0000ff}{\color[HTML]{FFFFFF} 2322 } & \cellcolor[HTML]{FF9595} 84 & \cellcolor[HTML]{FFE9E9} 18 & 0 & \cellcolor[HTML]{FFFFFF} 1 & \cellcolor[HTML]{FFFFFF} 1 & 0 & 0 & \cellcolor[HTML]{E3DCCC} 2471 \\ 
   \hhline{~|-|-|-|-|-|-|-|-|-|-|-|-|-|-|-|-|}\noalign{\vskip.2pt}
 & \cellcolor[HTML]{E3DCCC} 9 & 0 & 0 & \cellcolor[HTML]{FFFFFF} 1 & \cellcolor[HTML]{FFFFFF} 1 & \cellcolor[HTML]{FFFDFD} 2 & \cellcolor[HTML]{FF6666} 121 & \cellcolor[HTML]{FF1D1D} 178 & \cellcolor[HTML]{0000ff}{\color[HTML]{FFFFFF} 8592 } & \cellcolor[HTML]{FF0000} 201 & \cellcolor[HTML]{FF8D8D} 90 & \cellcolor[HTML]{FFD6D6} 33 & \cellcolor[HTML]{FFF9F9} 5 & 0 & \cellcolor[HTML]{FFFFFF} 1 & \cellcolor[HTML]{E3DCCC} 9225 \\ 
   \hhline{~|-|-|-|-|-|-|-|-|-|-|-|-|-|-|-|-|}\noalign{\vskip.2pt}
 & \cellcolor[HTML]{E3DCCC} 10 & 0 & 0 & 0 & 0 & 0 & \cellcolor[HTML]{FFF2F2} 11 & \cellcolor[HTML]{FFE1E1} 24 & \cellcolor[HTML]{FF4646} 146 & \cellcolor[HTML]{0000ff}{\color[HTML]{FFFFFF} 4624 } & \cellcolor[HTML]{FF6262} 124 & \cellcolor[HTML]{FFC6C6} 45 & \cellcolor[HTML]{FFF8F8} 6 & 0 & 0 & \cellcolor[HTML]{E3DCCC} 4980 \\ 
   \hhline{~|-|-|-|-|-|-|-|-|-|-|-|-|-|-|-|-|}\noalign{\vskip.2pt}
 & \cellcolor[HTML]{E3DCCC} 11 & 0 & 0 & 0 & 0 & 0 & \cellcolor[HTML]{FFFDFD} 2 & \cellcolor[HTML]{FFF6F6} 8 & \cellcolor[HTML]{FFD1D1} 37 & \cellcolor[HTML]{FF6666} 121 & \cellcolor[HTML]{0000ff}{\color[HTML]{FFFFFF} 5236 } & \cellcolor[HTML]{FF6464} 122 & \cellcolor[HTML]{FFDADA} 30 & \cellcolor[HTML]{FFFDFD} 2 & \cellcolor[HTML]{FFFFFF} 1 & \cellcolor[HTML]{E3DCCC} 5559 \\ 
   \hhline{~|-|-|-|-|-|-|-|-|-|-|-|-|-|-|-|-|}\noalign{\vskip.2pt}
 & \cellcolor[HTML]{E3DCCC} 12 & 0 & 0 & 0 & 0 & \cellcolor[HTML]{FFFFFF} 1 & \cellcolor[HTML]{FFF9F9} 5 & \cellcolor[HTML]{FFFDFD} 2 & \cellcolor[HTML]{FFE8E8} 19 & \cellcolor[HTML]{FFBCBC} 53 & \cellcolor[HTML]{FF6B6B} 117 & \cellcolor[HTML]{0000ff}{\color[HTML]{FFFFFF} 5708 } & \cellcolor[HTML]{FFCCCC} 41 & \cellcolor[HTML]{FFEFEF} 13 & \cellcolor[HTML]{FFF8F8} 6 & \cellcolor[HTML]{E3DCCC} 5965 \\ 
   \hhline{~|-|-|-|-|-|-|-|-|-|-|-|-|-|-|-|-|}\noalign{\vskip.2pt}
 & \cellcolor[HTML]{E3DCCC} 13 & 0 & 0 & 0 & 0 & 0 & 0 & 0 & \cellcolor[HTML]{FFFDFD} 2 & 0 & \cellcolor[HTML]{FFF0F0} 12 & \cellcolor[HTML]{FFEBEB} 16 & \cellcolor[HTML]{0000ff}{\color[HTML]{FFFFFF} 1642 } & \cellcolor[HTML]{FFEFEF} 13 & \cellcolor[HTML]{FFF8F8} 6 & \cellcolor[HTML]{E3DCCC} 1691 \\ 
   \hhline{~|-|-|-|-|-|-|-|-|-|-|-|-|-|-|-|-|}\noalign{\vskip.2pt}
 & \cellcolor[HTML]{E3DCCC} 14 & 0 & 0 & 0 & 0 & 0 & 0 & 0 & 0 & 0 & \cellcolor[HTML]{FFFFFF} 1 & \cellcolor[HTML]{FFF8F8} 6 & \cellcolor[HTML]{FFEFEF} 13 & \cellcolor[HTML]{0000ff}{\color[HTML]{FFFFFF} 890 } & \cellcolor[HTML]{FFF9F9} 5 & \cellcolor[HTML]{E3DCCC} 915 \\ 
   \hhline{~|-|-|-|-|-|-|-|-|-|-|-|-|-|-|-|-|}\noalign{\vskip.2pt}
 & \cellcolor[HTML]{E3DCCC} 15 & 0 & 0 & 0 & 0 & 0 & 0 & 0 & 0 & \cellcolor[HTML]{FFFFFF} 1 & 0 & 0 & \cellcolor[HTML]{FFFFFF} 1 & \cellcolor[HTML]{FFEFEF} 13 & \cellcolor[HTML]{0000ff}{\color[HTML]{FFFFFF} 2113 } & \cellcolor[HTML]{E3DCCC} 2128 \\ 
   \hhline{~|-|-|-|-|-|-|-|-|-|-|-|-|-|-|-|-|}\noalign{\vskip.2pt}
 & \cellcolor[HTML]{E3DCCC} Subtotal & \cellcolor[HTML]{E3DCCC} 25 & \cellcolor[HTML]{E3DCCC} 61 & \cellcolor[HTML]{E3DCCC} 367 & \cellcolor[HTML]{E3DCCC} 314 & \cellcolor[HTML]{E3DCCC} 1034 & \cellcolor[HTML]{E3DCCC} 2792 & \cellcolor[HTML]{E3DCCC} 2587 & \cellcolor[HTML]{E3DCCC} 8932 & \cellcolor[HTML]{E3DCCC} 5020 & \cellcolor[HTML]{E3DCCC} 5583 & \cellcolor[HTML]{E3DCCC} 5932 & \cellcolor[HTML]{E3DCCC} 1739 & \cellcolor[HTML]{E3DCCC} 931 & \cellcolor[HTML]{E3DCCC} 2132 & \cellcolor[HTML]{E3DCCC} 37449 \\ 
   \hhline{~|-|-|-|-|-|-|-|-|-|-|-|-|-|-|-|-|}\noalign{\vskip.2pt}
\end{tabular}
}
\end{sidewaystable}

\newpage

\begin{table}[htb]
\caption{Managers Heterogeneity}
\label{tab:managers}
\begin{threeparttable}
\begin{tabular}{lccc}
\hline
                    & \multicolumn{3}{c}{Results with the scorecard rating as predictor}    \\
                    & Number of Managers     & Error Increment +     & Error Increment -    \\ \hline
Significant at 10\% & 23                     & 2.51                  &                      \\
Significant at 5\%  & 14                     & 2.92                  &                      \\
Significant at 1\%  & 4                      & 3.76                  &                      \\
Total               & 328                    &                       &                      \\ \hline
                    &                        &                       &                      \\
                    & \multicolumn{3}{c}{Results without the scorecard rating as predictor} \\ \hline
Significant at 10\% & 30                     & 2.27                  & 0.36                 \\
Significant at 5\%  & 18                     & 2.54                  & 0.37                 \\
Significant at 1\%  & 3                      & 2.95                  &                      \\
Total               & 328                    &                       &                      \\ \hline
\end{tabular}%
\begin{tablenotes}
\small
\item The table shows the results for a regression of the demeaned absolute error from the cross-validation on manager dummies. The first column shows the significance level, the second column shows how many managers had absolute errors significantly different from the average and the remaining two columns show the mean absolute error of the significant managers divided by the mean absolute error of all managers for managers which had significantly bigger errors than the average and manager that had significantly smaller errors than the average. Empty cells indicate that there were no managers in the respective category. 
\end{tablenotes}
\end{threeparttable}
\end{table}

\begin{table}[htb]
\caption{Industry Heterogeneity}
\label{tab:industry}
\begin{threeparttable}
\begin{tabular}{lcc}
\hline
\multicolumn{3}{c}{Results with the scorecard rating as predictor}                           \\
NAICS Code & Industry Name                                                 & Error Increment \\ \hline
23         & Construction *                                                & 0.73            \\
31         & Manufacturing *                                               & 1.25            \\
44         & Retail Trade **                                               & 0.74            \\
52         & Finance and Insurance *                                       & 1.24            \\
62         & Health Care and Social Assistance *                           & 1.18            \\
72         & Accommodation and Food Services **                            & 0.44            \\ \hline
           &                                                               &                 \\
\multicolumn{3}{c}{Results without the scorecard rating as predictor}                        \\ \hline
44         & Retail Trade *                                                & 0.8             \\
52         & Finance and Insurance ***                                     & 1.34            \\
56         & Adm. and Supp., Waste Manag. and Remed. Serv. ** & 0.77            \\
62         & Health Care and Social Assistance **                          & 1.19            \\ \hline
\multicolumn{3}{l}{* Significant at 10\%, ** significant at 5\%, *** significant at 1\%}    
\end{tabular}%
\begin{tablenotes}
\small
\item The table shows the results for a regression of the demeaned absolute error from the cross-validation on industry dummies. The number if industries is 24. The last column shows the mean absolute error of the respective industry divided by the mean absolute error of all industries. 
\end{tablenotes}
\end{threeparttable}
\end{table}

\begin{table}
\begin{threeparttable}
\caption{Random Forest Cross Validation Prediction Results by Year }
\label{tab:cv_year}
\resizebox{\textwidth}{!}{%
\begin{tabular}{lccccc}
\hline
\multicolumn{6}{c}{Results with the scorecard rating as predictor}                \\
year     & Samp. Size & Accuracy & AccuracyLower & AccuracyUpper & AccuracyPValue \\ \hline
2010 & 4165   & 0.912    & 0.903         & 0.920         & 0.000          \\
2011 & 4149   & 0.907    & 0.898         & 0.916         & 0.000          \\
2012 & 4161   & 0.921    & 0.912         & 0.929         & 0.000          \\
2013 & 4161   & 0.931    & 0.923         & 0.938         & 0.000          \\
2014 & 4155   & 0.945    & 0.938         & 0.952         & 0.000          \\
2015 & 4162   & 0.932    & 0.924         & 0.940         & 0.000          \\
2016 & 4164   & 0.937    & 0.930         & 0.944         & 0.000          \\
2017 & 4171   & 0.961    & 0.955         & 0.967         & 0.000          \\
2018 & 4161   & 0.964    & 0.958         & 0.970         & 0.000          \\ \hline
         &            &          &               &               &                \\
\multicolumn{6}{c}{Results without the scorecard rating as predictor}             \\ \hline
2010 & 4165   & 0.895    & 0.885         & 0.904         & 0.000          \\
2011 & 4149   & 0.883    & 0.873         & 0.893         & 0.000          \\
2012 & 4161   & 0.904    & 0.895         & 0.913         & 0.000          \\
2013 & 4161   & 0.913    & 0.905         & 0.922         & 0.000          \\
2014 & 4155   & 0.933    & 0.925         & 0.940         & 0.000          \\
2015 & 4162   & 0.921    & 0.912         & 0.929         & 0.000          \\
2016 & 4164   & 0.915    & 0.906         & 0.923         & 0.000          \\
2017 & 4171   & 0.945    & 0.937         & 0.952         & 0.000          \\
2018 & 4161   & 0.958    & 0.952         & 0.964         & 0.000   \\ \hline      
\end{tabular}%
}%
\begin{tablenotes}
\small
\item The table shows the results of a 5-fold cross-validation for the  Random Forest for each year in the sample. The Accuracy p-value comes from a one-sided test \citep{kuhn2008building} to see if the accuracy is better than the "no information rate", which is the largest class percentage in the data (23.85\%). The confidence interval is of 95\% and it comes from a binomial test. The first set of rows show the results for models where we included the Scorecard rating as a predictive variable and the second set rows show the results for models where we did not include the Scorecard rating as a predictive variable. 
\end{tablenotes}
\end{threeparttable}
\end{table}

\end{document}